\documentclass[useAMS,usenatbib]{mn2e}	
\usepackage{graphicx}
\usepackage{subfigure}
\usepackage{verbatim}

\title[The early triggering of quasar activity]{Clear evidence for the early triggering of a luminous quasar-like AGN in a major, gas rich merger}
\author[P.~S.~ Bessiere et al.]
{\parbox{\textwidth}{P.~S.~ Bessiere$^{1}$\thanks{E-mail:p.bessiere@sheffield.ac.uk},
C.~N.~Tadhunter$^{1}$,	
C.~Ramos Almeida$^{2,3}$,
\&
M.~Villar Mart\'{i}n$^{4}$}\vspace{0.4cm}\\
\parbox{\textwidth}{$^{1}$ Department of Physics and Astronomy, University of Sheffield, Sheffield, S3 7RH, UK\\
${^2}$ Instituto de Astrof\'\i sica de Canarias (IAC), C/V\'ia L\'actea, s/n, E-38205, La Laguna, Tenerife, Spain\\
${^3}$ Departamento de Astrof\'\i sica, Universidad de La Laguna, E-38205, La Laguna, Tenerife, Spain\\
${^4}$ Centro de Astrobiolog\'{i}a (INTA-CSIC), Carretera de Ajalvier, km 4, 28850 Torrej\'{o}n de Ardoz, Madrid, Spain
}}

\begin{document}
\bibliographystyle{mn2e}
\date{}
\pagerange{\pageref{firstpage}--\pageref{lastpage}} \pubyear{2013}
\maketitle
\label{firstpage}

\begin{abstract}
We present deep, intermediate resolution, long slit Gemini GMOS-S optical spectra of the SDSS type II quasar J002531-104022, which is a highly disturbed system currently undergoing a major merger event. We use these data to model the ages and reddenings of the stellar populations in three distinct spatial regions and find a remarkable uniformity in the properties of the young stellar populations (YSP) that dominate the optical spectra. The YSPs are all found to have relatively young ages ($t_{ysp}< 40$ Myr), strongly implying that the latest episode of star formation and quasar activity have been triggered quasi-simultaneously. The lack of reddening deduced from both continuum modelling and the measured Balmer decrements (E(B-V)$<0.3$) suggests that starburst and/or AGN induced outflows have already effectively removed a substantial proportion of the gas and dust from the central region. These findings starkly contrast with model predictions which suggest an offset of a few 100 Myr between the peak of merger induced star formation and the emergence of the optically visible quasar activity. Based on our stellar population fits we also show that the total stellar mass is in the range $(4 - 17) \times 10^{10}~M_{\sun}$, lower than typically found for quasar host galaxies. 
\end{abstract}


\section{Introduction}
\label{intro}
There is a growing acceptance that active galactic nuclei (AGN) have played a fundamental role in the evolution of galaxies. This is mainly due to the discovery of strong correlations between the masses of the black holes (BH) which are thought to reside at the heart of all massive galaxies \citep{magorrian98} and many of the properties of the galaxy bulges (e.g. \citealt{ferrarese00, gebhardt00}). Despite this, many unanswered questions still remain about how and when AGN are triggered.

It has been suggested that, in the case of powerful quasars, major mergers between gas rich galaxies are often the triggering mechanism \citep{toomre72,heckman86,hopkins06}. This is largely based on the idea that such gas-rich mergers are capable of providing the necessary torques to deliver large reservoirs of gas to the central kpc, where they might trigger quasar activity. Such events would leave observable signatures in the form of double nuclei, tidal tails and other tidal features. Observational studies aimed at detecting these features have provided some evidence for this merger driven scenario, with a high rate of detection (up to 95\% in powerful radio galaxies) of tidal features associated with luminous AGN host galaxies \citep{bennert08,ramos11,bessiere12, villar12}. Notably, when compared with a sample of quiescent galaxies matched in redshift and absolute magnitude, both \cite{ramos12} and \cite{bessiere12} find that the surface brightnesses of tidal features associated with the luminous AGN are up to two magnitudes brighter, perhaps suggesting that these mergers are more recent or more gas rich. These studies also show that the quasar activity can be triggered in a variety of merger phases, including the pre-coalesence phase. 

Quasar activity is not the only consequence of such merger events. Simulations suggest that the cold gas injected into the nuclear region by the merger also provides the conditions for a massive burst of star formation \citep{dimatteo05,springel05}.  If, as simulations suggest, the triggering of these starbursts and the onset of AGN activity are both the result of the same merger, then determination of the ages of starbursts in a large enough sample of AGN host galaxies has the potential to allow us to understand the timing of the triggering of the AGN, relative to the main merger event. Thus, we can track the evolution of mergers from the initial first pass, through the triggering of the quasar activity, and on to the final coalescence of the BHs.

Dating the stellar populations also allows us to test evolutionary scenarios, such as that put forward by \citet{sanders88}, in which mergers between gas rich galaxies lead to ultra-luminous infrared galaxies (ULIRGs), where the IR luminosity is initially driven by rapid star formation. Later in the merger, at the point of coalescence of the two BHs, the AGN begins to dominate the IR luminosity, eventually driving out its cocoon of dust and gas, and allowing it to become detectable in the optical  (e.g. \citealt{dimatteo05,springel05,hopkins06}). In this scenario, we would expect that the peak of star formation would occur while the AGN is still deeply buried, and therefore, by the time the AGN becomes visible in the optical, the stellar population would be detected as an ageing starburst of several 100 Myr. Although, as previously mentioned, morphological studies have already shown this scenario to be over simplistic, they provide little detailed information about the nature of the merger and the timing of the quasar relative to the merger-induced starburst.

A number of studies focussed on characterising the stellar populations of AGN host galaxies have detected relatively old post-starburst populations with ages 100 Myr to a few Gyr \citep{tadhunter05,holt07,wills08, wild10,canalizo13} which, if we assume a quasar lifetime of 1 -- 100 Myr \citep{martini01}, implies a substantial delay between the merger-induced starburst and the quasar being triggered/becoming visible, others have found clear evidence of quasi-simultaneous triggering of the luminous quasar and starburst \citep{heckman97,canalizo00,brotherton02,davies07,wills08,tadhunter11}. Therefore, the matter of whether luminous quasars are triggered at the peak of star formation in major, gas-rich mergers remains ambiguous.

\begin{figure}
\centering
\subfigure[]{
\includegraphics[]{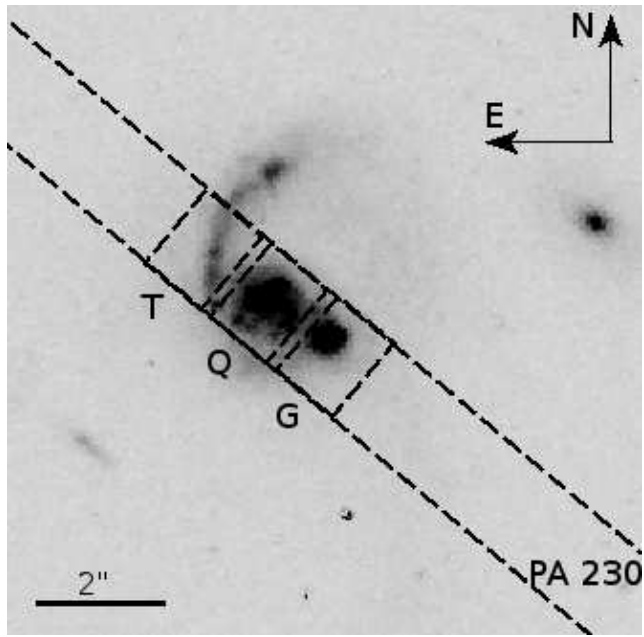}
\label{slit}}
\subfigure[]{
\includegraphics[trim = 10mm 0mm 10mm 0mm, scale = 0.5]{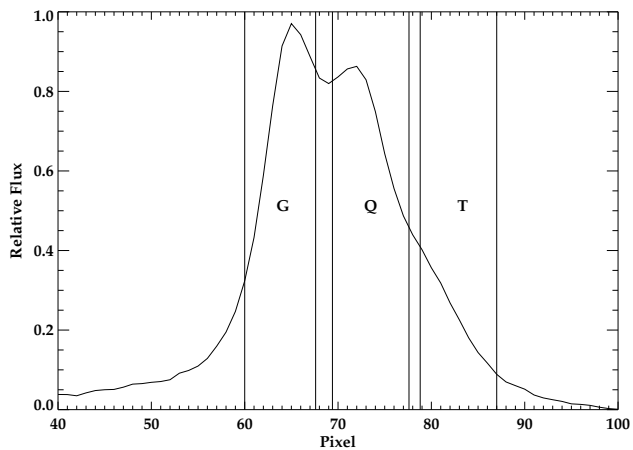}
\label{apertures}}

\caption{(a) HST image of J0025-10 showing the position of the 1.5 \arcsec slit used for our long slit spectroscopy. The image also shows the approximate spatial locations of the three 5 kpc apertures used in this analysis. (b) A spatial profile of the 2D spectrum of J0025-10 in the observed wavelength range 5370 -- 5553 \AA\, indicating the regions from which the three apertures were extracted. }
\label{loc}
\end{figure}

In order to clarify the sequence and timing of the events surrounding the triggering of luminous AGN, we have undertaken a campaign of optical spectroscopy of a sample of type II quasars (defined as having $L_{[OIII]} > 10^{8.5} L_{\sun}$, \citealt{zakamska03}), for which a full morphological study, based on deep Gemini GMOS-S $r^\prime$ imaging, has already been presented in \citet{bessiere12}. Here we present long-slit spectroscopic observations of one of the objects in our sample -- J0025-10 -- which shows some remarkable properties that make it worthy of note in the context of the evolutionary scenarios described above. The spectra are used to determine the ages, reddening and masses of the stellar populations in the host galaxy. Figure \ref{slit}\footnote{This image of J0025-10 is taken from the Hubble Legacy Archive. It was taken as part of proposal no. 10880 (P.I. Schmitt) using ACS/WFC with the F775W filter}shows that the J0025-10 host galaxy, at a redshift of z=0.303, is undergoing a major merger event, with the two distinct nuclei at a projected separation of $\sim 5$ kpc \citep{bessiere12}, and a projected velocity shift of only $\Delta v = -20 \pm 20$ km s$^{-1}$ \citep{villar11} between them. There are also two distinct tidal tails, the brighter of which is clearly visible curving to the north of the nucleus in Figure \ref{slit}. CO measurements made by \citet{villar13} reinforce the idea that this is a gas rich merger, showing that J0025-10 contains a total of $(6 \pm 1) \times 10^9$ M$_{\sun}$ of molecular gas in two reservoirs, with 60\% in the central region and 40\% associated with the northern tidal tail. They also determine that the system can be classified as a (U)LIRG with an infrared luminosity of $L_{IR} = (1.1 \pm 0.3) \times 10^{12}~L_{\sun}$

Throughout this paper, we assume a cosmology of $H_0 = 70$ km s$^{-1}$ Mpc$^{-1}$, $\Omega_M = 0.3$ and $\Lambda = 0.7$ which gives a scale of 4.48 kpc arcsec$^{-1}$ as the redshift of this object.

\section{Observations and data reduction}

Long-slit spectra were taken for J0025-10 on 30/11/2010, in queue mode, using the Gemini Multi-Object Spectrograph (GMOS) mounted on the 8.1m Gemini South telescope at Cerro Pach{\'o}n, Chile, as part of a larger Gemini program, aimed at characterising the stellar populations of the host galaxies of type II quasars (Bessiere et al. 2013, in prep). Due to the absence of an atmospheric dispersion compensator on GMOS-S, the object was observed at low airmass ($<1.1$) in order to minimise the effects of differential refraction. Our observations were taken at PA 230 degrees, in photometric conditions, with good seeing (0.8 \arcsec~ FWHM) and consist of $4 \times 675$ s exposures using the B600 grating with 600 groves mm$^{-1}$ blazed at 461 nm, and $3\times 400$ s exposures using the R400 grating with 400 groves mm$^{-1}$ blazed at 764 nm. The exposures in the two wavelength ranges were interleaved in order to mitigate any changes in the atmospheric conditions throughout the observational run. 

The data were reduced using {\sevensize IRAF\footnote{IRAF is distributed by the National Optical Astronomy Observatory, which is operated by the Association of Universities for Research in Astronomy (AURA) under cooperative agreement with the National Science Foundation.}} and {\sevensize STARLINK FIGARO\footnote{The authors acknowledge the data analysis facilities
provided by the Starlink Project, which is run by CCLRC
on behalf of PPARC.}} packages, following the standard steps of bias subtraction, flat fielding, wavelength calibration, flux calibration, and corrections for spatial distortions; the useful observed wavelength range is 3940 -- 9140 \AA, although wavelengths longer than 8200 \AA\ are strongly affected by fringes.  Accurate flux calibrations are essential when modelling stellar populations, therefore as part of our Gemini program several standard stars were observed throughout the semester (2010B), allowing us to achieve a flux calibration accuracy of $\pm 5\%$. There is also a large region of overlap between the red and blue parts of the spectra, allowing us to further check that the calibrations in the two parts matches well, and in this case there is less than a 1\% difference in flux in the overlapping region. Using a 1.5 arcsec slit, we achieved a spectral resolution of 7.2 \AA\ in the blue and 11.4 \AA\ in the red wavelength ranges. 

In addition to the spectroscopic data presented here, we also make use of deep Gemini GMOS-S $r^\prime$-band imaging data present in \citet{bessiere12}, which were reduced using the standard Gemini GMOS {\sevensize IRAF} packages. The $r^\prime$ magnitude of J0025-10 is $r_{AB}$=18.40 mag, measured within a 30 kpc diameter aperture and corrected for Galactic extinction, but with no k-correction applied.

\section{Results}

The analysis presented here is based on 1D spectra extracted from three different components of the system (Figure \ref{apertures}),  where each aperture covers a projected region of $1.5 \times 1.1\arcsec$ ($6.7 \times 5.0$ kpc). Apertures Q and G are centred on the nuclei of the quasar host galaxy and the merging companion galaxy respectively, whilst aperture T is centred on the tidal tail associated with the quasar host galaxy. The spectra have been corrected for Galactic reddening (E(B - V) = 0.03) as derived from the dust map of \citet*{schlegel98} using the extinction law of \citet{cardelli89}. 

\begin{table}
\caption{The H$\gamma/$H$\beta$ and H$\delta/$H$\beta$ ratios measured in each of the three apertures, following subtraction of a continuum model. Column one denotes which aperture the measurement applies to, columns two and three give the value of the H$\gamma/$H$\beta$ and H$\delta/$H$\beta$ ratio respectively.}
  \label{reddening}
  \begin{tabular}{l c c}
  \hline
  Aperture	&	H$\gamma/$H$\beta$		& H$\delta/$H$\beta$\\
  \hline   
   Q		&		$0.41\pm 0.03$		&	$0.22\pm0.02$		 \\
   G		& 		$0.47 \pm 0.04 $	& 	$0.25\pm0.02$\\
	 T		&		$0.51\pm0.04$		& 	$0.29\pm0.03$\\	
   \hline
  
  \end{tabular}
  \end{table}

\begin{figure*}
\centering
\includegraphics[trim = 100mm 0mm 100mm 0mm, scale = 1]{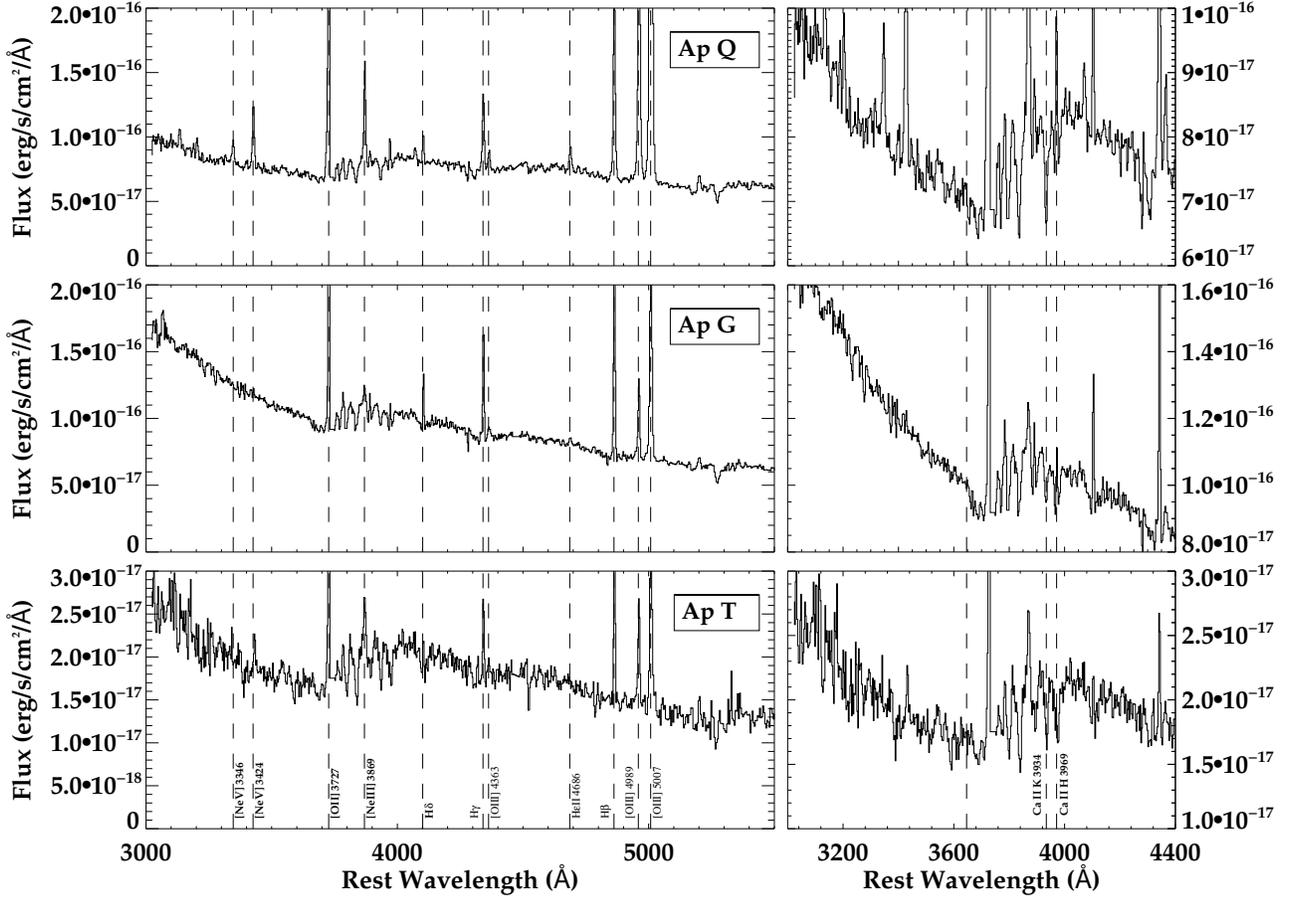}
\caption{The 1D spectra extracted from the three distinct spatial regions of J0025-10. Each spectrum is shown in a separate panel on the left, with the top panel showing the quasar host galaxy component (Q), the middle panel showing the companion galaxy component (G), and the bottom panel showing the spectrum for the northern tidal tail (T). The most significant emission lines are highlighted. The right hand panels show a zoom of the short wavelength end of the spectrum of same spatial region with prominent features highlighted. Region G shows the detection of the Balmer edge in emission which indicates the presence of a significant population of young stars and also demonstrates the importance of an accurate nebular subtraction when modelling stellar populations. }
\label{spectra}
\end{figure*}

Figure \ref{slit} shows the spatial location of each aperture, while Figure \ref{apertures} shows the location of the apertures plotted on a spatial profile of the 2D spectrum.  The extracted 1D spectra are shown in Figure \ref{spectra}, and in each of the apertures the presence of the higher order Balmer stellar absorption lines is clearly visible along with the Balmer break. These features alone indicate that significant young stellar populations must be present.

\subsection{Continuum Modelling}

In order to model the stellar continuum, we have used a purpose written {\sevensize IDL} routine ({\sevensize CONFIT}; \citealt{robinson00, robinson01}), which attempts to fit up to three components to the observed spectrum. These three components can comprise of up to two different stellar populations and, in addition, a power-law component which has the form $F_\lambda \propto \lambda^{\alpha}$ ($-15<  \alpha < 15 $). The routine fits the required number of components to the observed spectra using various combinations of flux, to compute the reduced minimum $\chi^2$ value for each combination. We have modelled the spectra using 77 wavelength bins of 30\AA\  width below a rest-frame wavelength of 6000\AA\, and 60\AA\ width at longer wavelengths; the wider bins at longer wavelength are necessary due to strong sky residuals  and fringing. These bins are evenly distributed across the wavelength range, avoiding strong emission lines, atmospheric absorption, chip gaps and bad columns. We then selected a normalising bin of 4583--4613\AA, because this region is free of strong emission lines, and proceeded to fit up to three components to the spectra. 

Before fitting, we also subtracted the higher order Balmer lines and a nebular continuum using the procedure outlined in \citet{dickson95} and \citet*{holt03}. The necessity of subtracting the nebular continuum is demonstrated by Figure \ref{spectra}, which shows clear evidence of the Balmer continuum edge in emission in the companion galaxy component. Measuring the contribution of the nebular continuum that we generate to the total flux just below the Balmer edge (3450--3550\AA\ ), we find that it contributes 2.7\%, 9.6\% and 8.8\% in apertures Q, G and T respectively. In order to determine whether it was necessary to apply a reddening correction to the subtracted nebular continuum, we measured the Balmer decrements from the stellar-continuum-subtracted spectrum, which was generated by subtracting the best fitting model from the data before the nebular continuum was accounted for. Unfortunately, because of the strong sky residuals and fringing above 6000 \AA\ (rest-frame), we were unable to accurately measure the H$\alpha$ flux, and therefore measured the H$\gamma/$H$\beta$ and H$\delta/$H$\beta$ ratios instead. The results, shown in Table \ref{reddening}, are consistent at the $2\sigma$ level with Case B recombination (H$\gamma/$H$\beta$ = 0.47, H$\delta/$H$\beta$ = 0.27; \citealt{osterbrock06}). This implies little to no intrinsic reddening in the various apertures, and therefore we apply no correction to the nebular component. Careful examination of the nebular subtracted spectra in the region of the Balmer edge confirmed the accuracy of our subtraction.

The stellar templates we use here were generated using Starburst 99 (Version 6.0.4) \footnote{http://www.stsci.edu/science/starburst99/docs/default.htm} \citep{leitherer99,vazquez05,leitherer10} assuming an instantaneous burst model with an initial mass of $10^6~M_{\sun}$, solar metalicity and a Kroupa IMF \citep{kroupa01}. Reddening in the range $ 0 \le E(B-V) \le 2.0$ was then applied to these template spectra using the extinction curve of \citet{calzetti00}. 

The strategy that we adopted for the stellar synthesis modelling was to use the minimum number of components required to produce an acceptable fit. In this way, we sought to minimise the issue of degeneracies inherent in fitting larger numbers of components. For each aperture we ran the model with a series of combinations of components, which represent plausible evolutionary scenarios for the host galaxies, and included the following.

\begin{enumerate}

\item An unreddend old stellar population (OSP) of age 8 Gyr combined with a young to intermediate stellar population (YSP/ISP) with age in the range 0.001 -- 5 Gyr and reddening in the range $ 0 \le E(B-V) \le 2.0$.
\item An intermediate age, unreddened stellar population (0.5--2 Gyr age) combined with a YSP/ISP with age in the range 0.001 -- 2 Gyr and reddening in the range $ 0 \le E(B-V) \le 2.0$. 
\item In addition, for the quasar host galaxy aperture (Q), we run fits (i) and (ii) including a power-law, which could be representative of a scattered AGN component \citep{tadhunter02}.
\end{enumerate}

\begin{figure*}
\centering
\begin{minipage}{190mm}
\subfigure[]{
\includegraphics[trim = 8mm 0mm 0mm 0mm, clip,width=6.2cm]{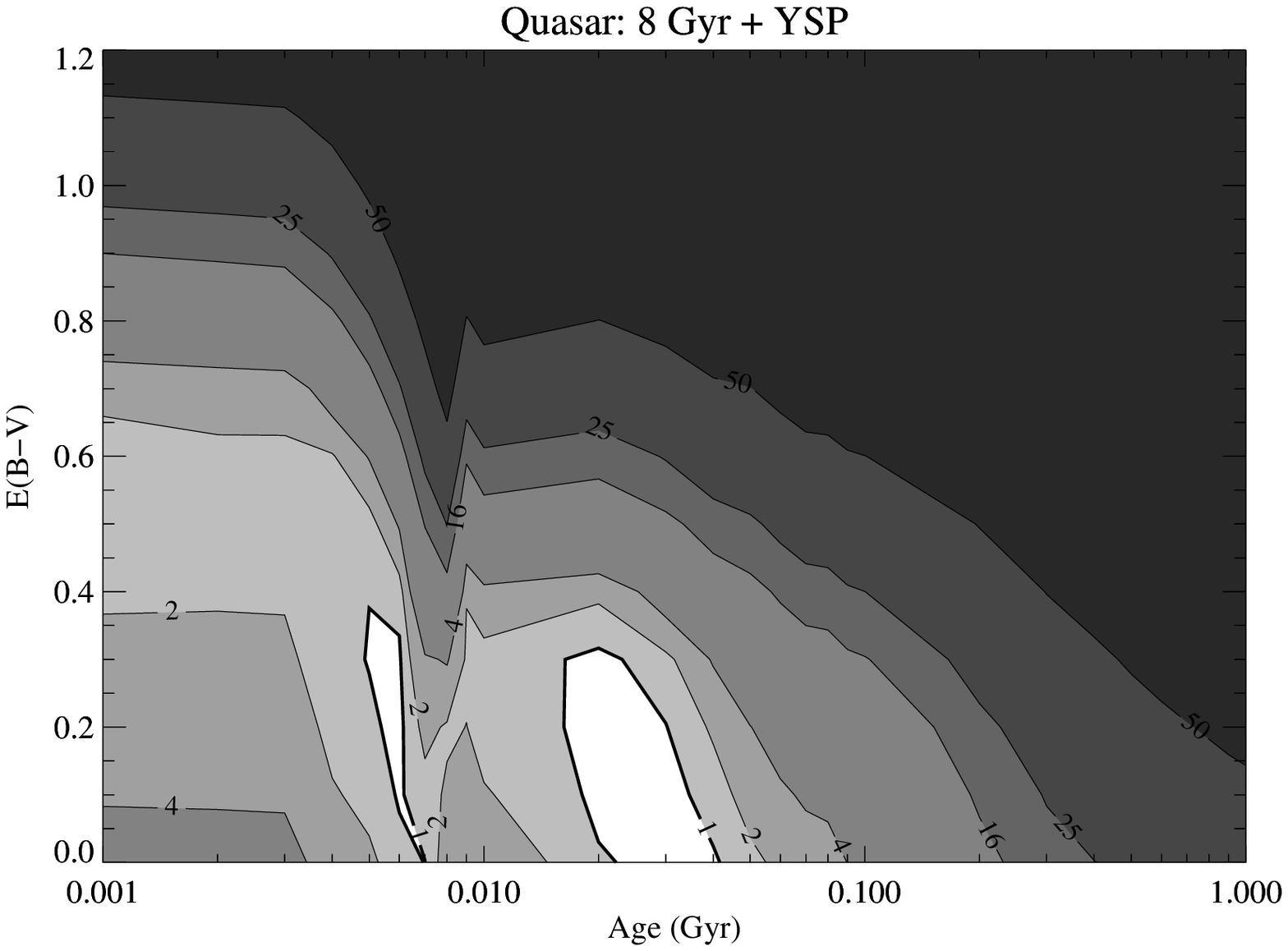}

\label{q8ysp}}
\subfigure[]{
\includegraphics[trim = 8mm 0mm 0mm 0mm, clip,width=6.2cm]{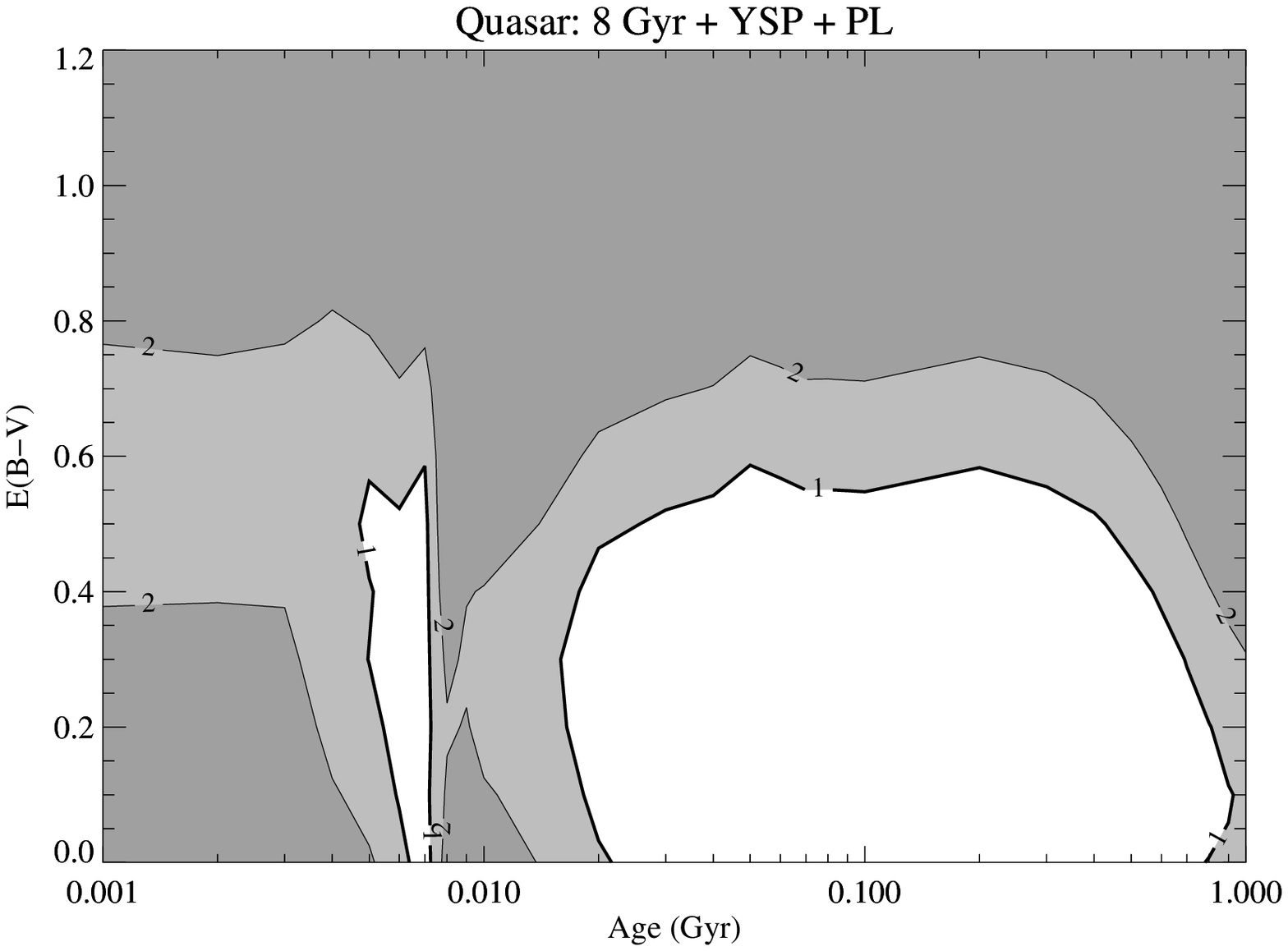}
\label{q8ysppl}}
\subfigure[]{
\includegraphics[trim = 8mm 0mm 0mm 0mm, clip,width=6.2cm]{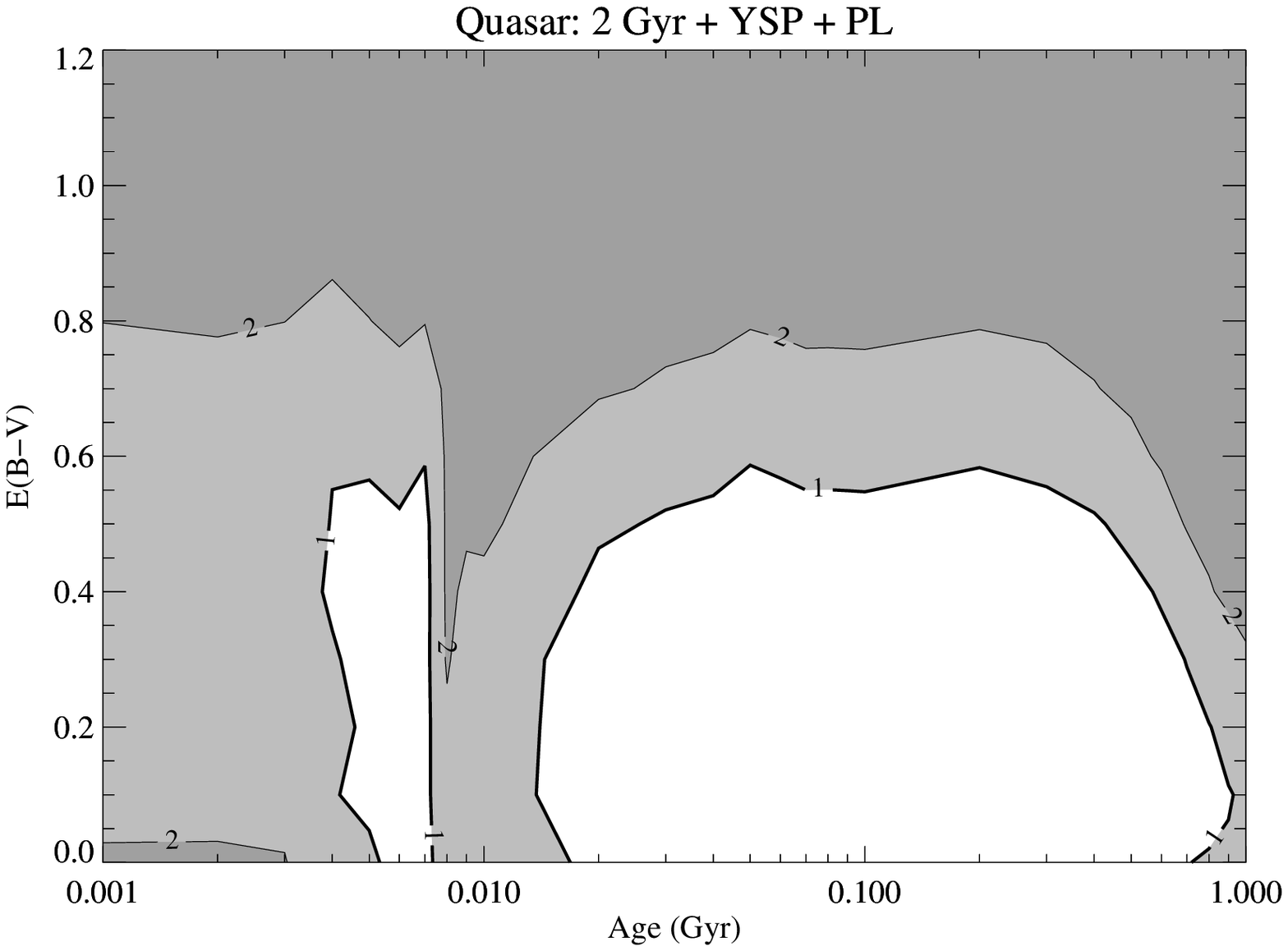}
\label{q2ysppl}}
\subfigure[]{
\includegraphics[trim = 8mm 0mm 0mm 0mm, clip,width=6.2cm]{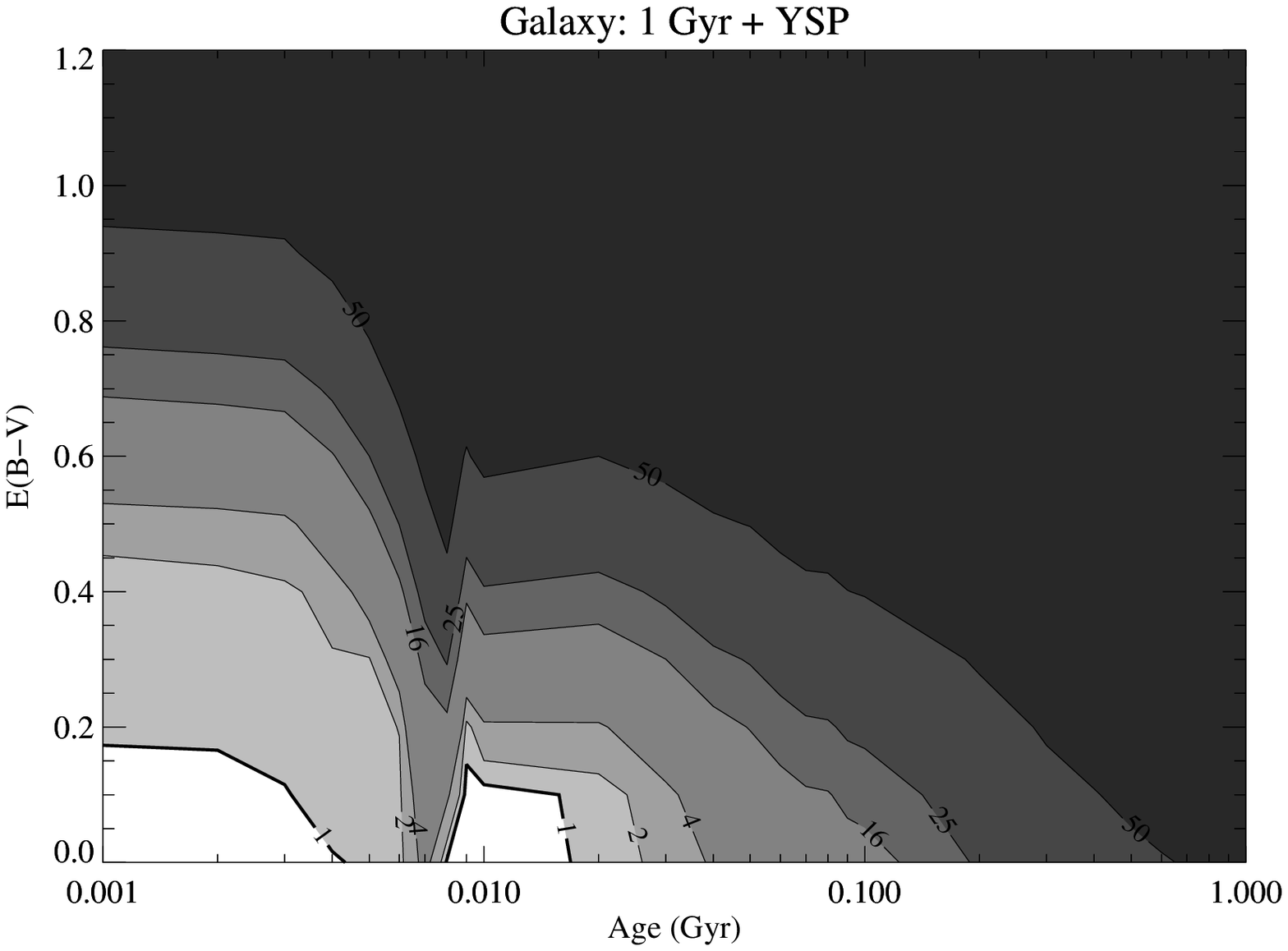}
\label{g1ysp}}
\subfigure[]{
\includegraphics[trim = 8mm 0mm 0mm 0mm, clip,width=6.2cm]{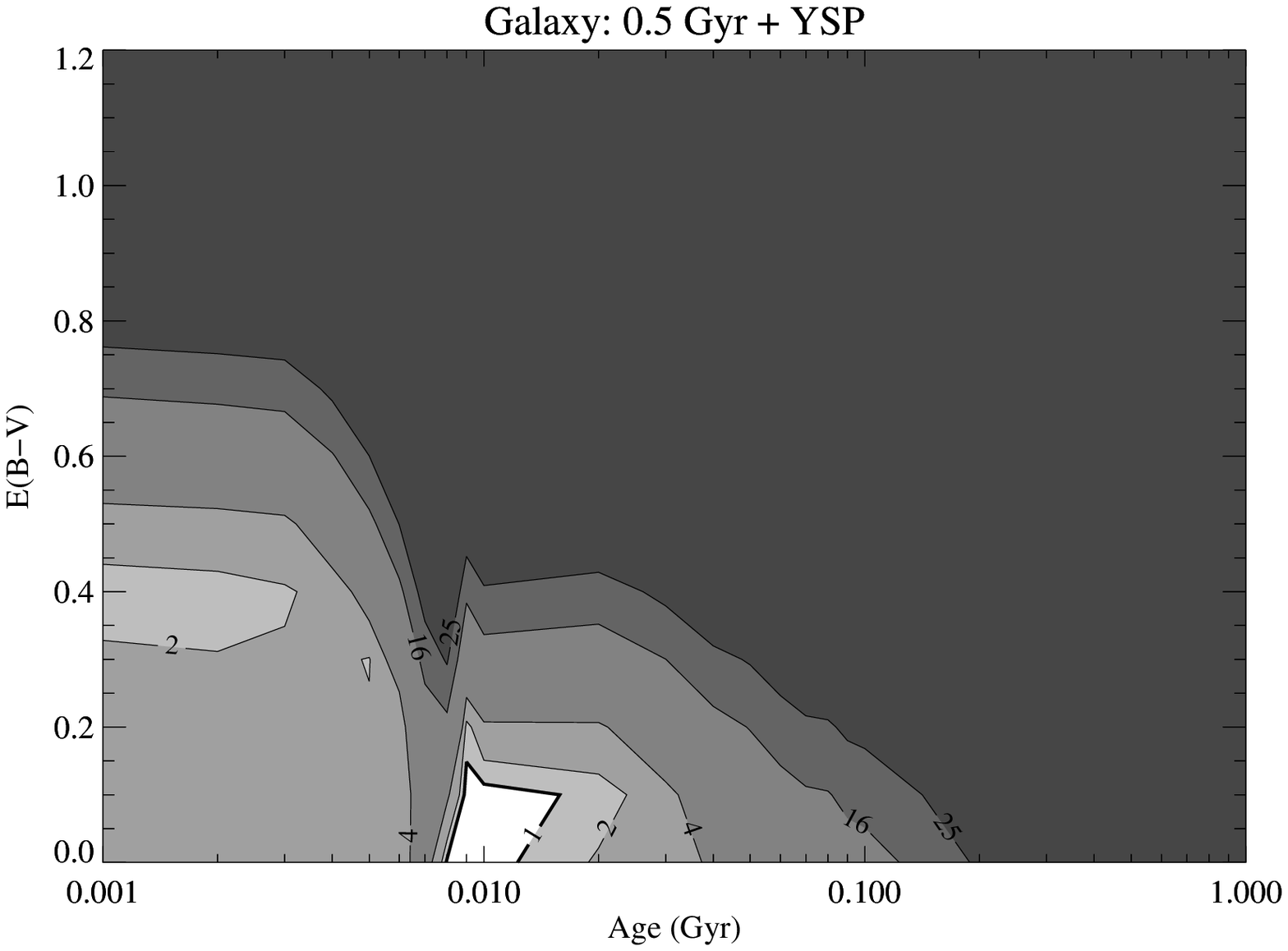}
\label{g0-5ysp}}
\subfigure[]{
\includegraphics[trim = 8mm 0mm 0mm 0mm, clip,width=6.2cm]{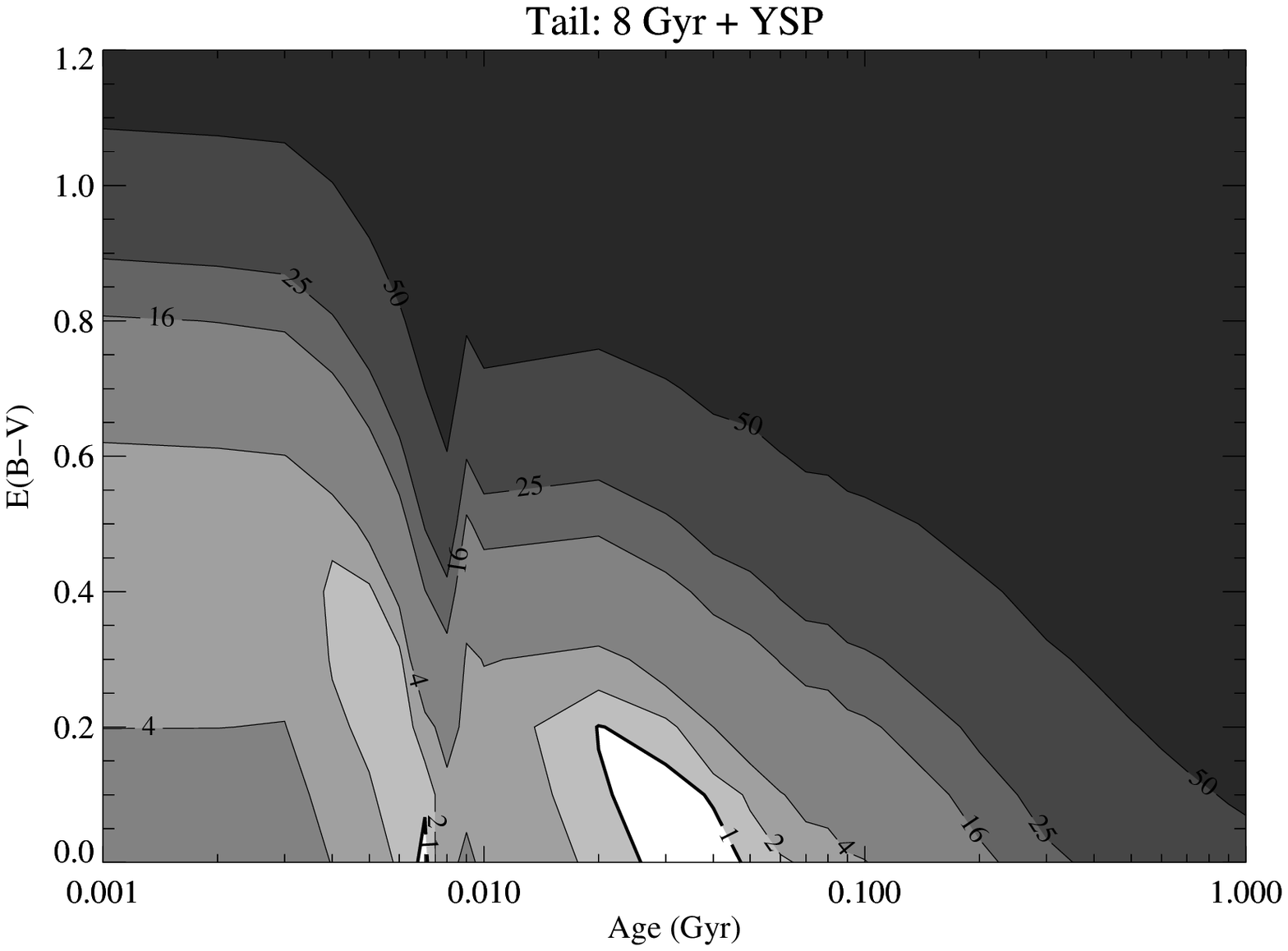}
\label{t8ysp}}
\subfigure[]{
\includegraphics[trim = 8mm 0mm 0mm 0mm, clip,width=6.2cm]{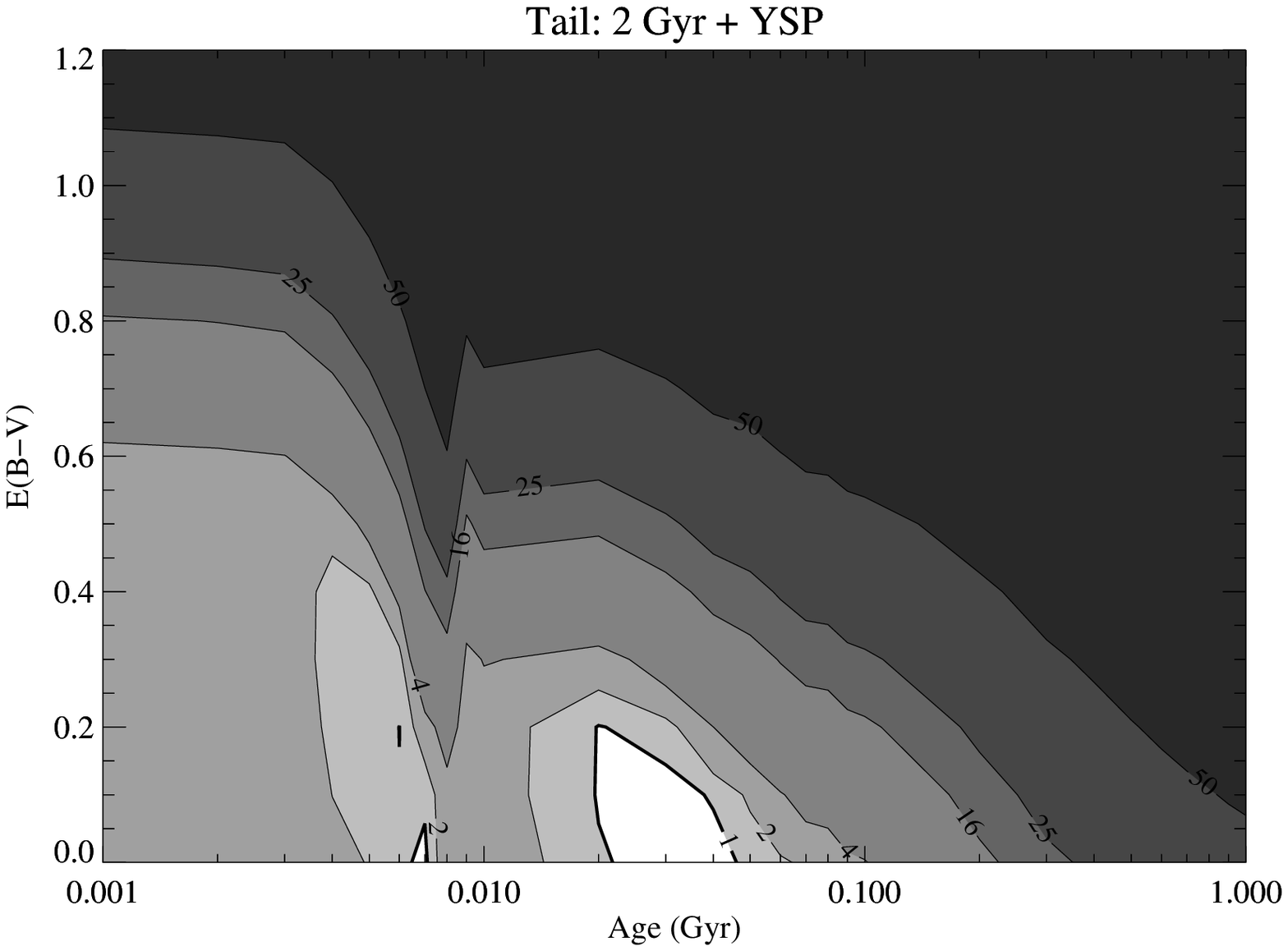}
\label{t2ysp}}
\subfigure[]{
\includegraphics[trim = 8mm 0mm 0mm 0mm, clip,width=6.2cm]{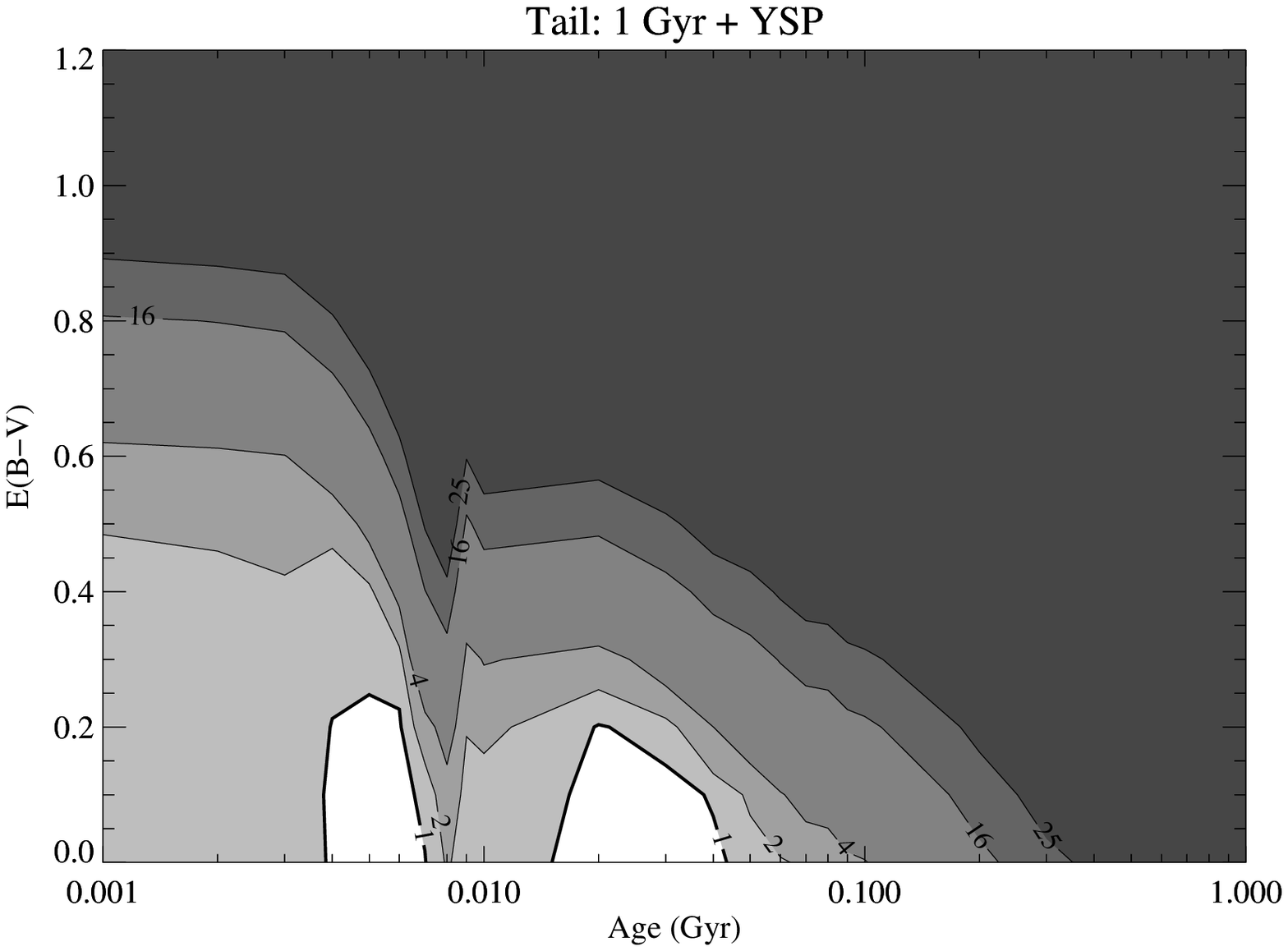}
\label{t1ysp}}
\subfigure[]{
\includegraphics[trim = 8mm 0mm 0mm 0mm, clip,width=6.2cm]{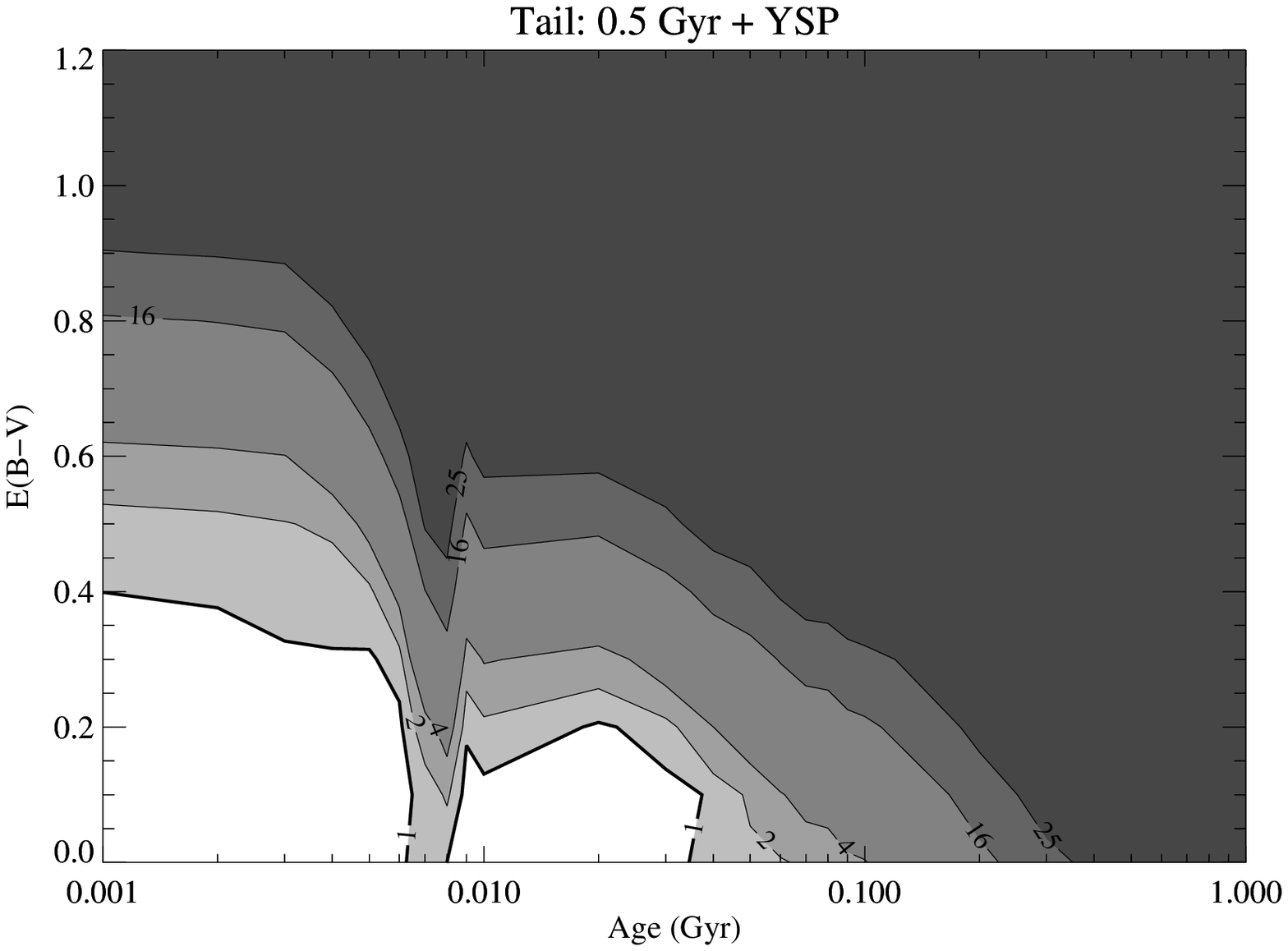}
\label{t0-5ysp}}
\caption{The reduced minimum $\chi^2$ contour plots for each of our successful {\sevensize CONFIT} runs. Each plot shows the region of $\chi^2$ ($\chi^2 < 1$) space in which possible acceptable fits may be found in white. The numbers give the reduced minimum $\chi^2$ value of the contour up to a maximum value of 50.}
\label{contour_plots}
\end{minipage}
\end{figure*}

\begin{figure}
\centering
\includegraphics[trim = 0mm 0mm 0mm 0mm, scale = 0.5]{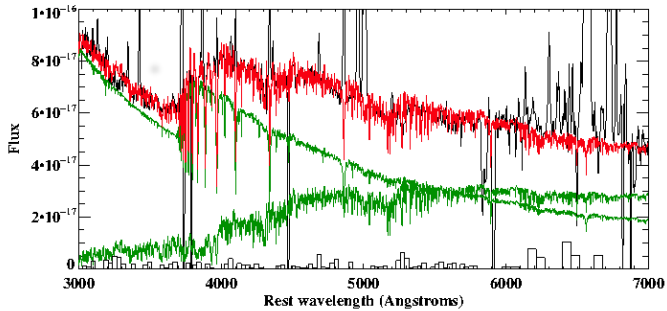}
\caption{An example of the overall fit to the spectrum, extracted from aperture Q, produced by the {\sevensize CONFIT} routine. The data is shown in black whilst the OSP and YSP models are in green. The combined overall fit to the data is shown in red and at the bottom, the residuals in the wavelength bins are shown.}
\label{mainfit}
\end{figure}

For each combination of components, we produced a contour plot (see Figure \ref{contour_plots}) which shows the contours of reduced minimum chi-squared ($\chi^2_{red}$) values for the various combinations of OSP/ISP and YSP with various ages and reddening, the results of which are summarised in Table \ref{fit_info}. Taking into consideration previous experience with such fits, any combination with $\chi^2_{red}	 < 1$ is deemed to be a possible acceptable fit (e.g. \citealt{tadhunter05,holt07}). We then further discriminated on the basis of visual inspection of how well some of the age-sensitive absorption lines, including the higher order Balmer lines and G band, were fitted. Figure \ref{mainfit}, shows an example of the overall fit produced by {\sevensize CONFIT} for aperture Q for a particular combination of stellar populations.

\begin{table*}
\centering
  \begin{minipage}{180mm}
  \caption{The results of the various combinations of stellar populations used in our {\sevensize CONFIT} runs. Column 1 gives the aperture that is being modelled. Columns 2 to 4 show the age of the OSP/ISP population used in the fit, whether a power law component was also included, and whether an acceptable fit was achieved. Columns 5 and 6 denote the age and reddening of the YSPs that allow an acceptable fit. Columns 7 to 9 give the percentage contribution to the total flux in the aperture of the OSP, YSP and, if it is included, the power law component respectively.}
  
  \label{fit_info}
  \begin{tabular}{l c c c l c c c c}
  \hline
   Aperture	& OSP/ISP	& Power law	& Acceptable	& YSP (Gyr)			& E(B-V)	& OSP/ISP 	& YSP flux \%	& Power Law \\
						& (Gyr)		& component	& fit ?		& 			&		& flux \%	& 		& flux \%	\\
   \hline

	Q	& 8			& No		& Yes		& 0.02--0.04		& 0--0.2	& 19 --37	& 63--79	& --\\	
	Q	& 2			& No		& No		& --			& --		& --		& --		& -- \\
	Q	& 1			& No		& No		& --			& --		& --		& --		& -- \\
	Q	& 0.5		& No		& No		& --			& --		& --		& --		& -- \\	
  Q	& 8			& Yes		& Yes		& 0.02 -- 0.3		& 0--0.5	& 0--39		& 34--79	& 4--39	\\
	Q	& 2			& Yes		& Yes		& 0.02--0.4		& 0--0.4	& 0--42		& 39--94	& 0--42	\\
	Q	& 1			& Yes		&	No		&	--					&	--			&	--			& --			&	\\
	Q	& 0.5		& Yes		& No		& --			& --		& --		& --		& --	\\

	G	& 8			& No		& No		& --			& --		& --		& --		& --	\\
	G	& 2			& No		& No		& --			& --		& --		& --		& --	\\
	G	& 1			& No		& Yes		& 0.009 --0.01			& 0--0.1		& 0--32	& 67--97	& --	\\
	G	& 0.5		& No		& Yes		& 0.009--0.01		& 0--0.1	& 0--37	& 65--96	&	\\	

	T	& 8			& No		& Yes		& 0.03--0.04		& 0--0.1	& 5--20		& 77--92	& --	\\
	T	& 2			& No		& Yes		& 0.03--0.04		& 0--0.1	& 6--24		& 73--91	& --	\\
	T	& 1			& No		& Yes		& 0.02--0.04 (0.006)	& 0--0.1 	& 7--44		& 53--90	& --	\\
	T	& 0.5		& No		& Yes		& 0.001--0.006 (0.009--0.03) 	&	0--0.3	& 7--66		& 34--90	& --	\\

  \hline
  \end{tabular}
  \end{minipage}       
\end{table*}

\subsection{The quasar host galaxy}

Figure \ref{q8ysp} shows the contour plot of $\chi^2_{red}$ values for the combination of an 8 Gyr OSP  with a YSP/ISP for the bulge of the quasar host galaxy. It is clear that there are two regions in which $\chi^2_{red} < 1$ (4-8 Myr and 20-40 Myr), however, although the fit to the general shape of the continuum with YSPs of age 4-8 Myr are reasonable, closer inspection of the fit in the region of the higher order Balmer lines allows us to disregard these as acceptable solutions. Therefore, if a power-law component is not included, then acceptable fits are achieved with an OSP of 8 Gyr, combined with a YSP component with age between 20 and 40 Myr, and low reddening in the range $0 \le E(B-V) \le 0.2 $. Table \ref{fit_info} shows that, for all combinations for which an acceptable fit is achieved, the flux is dominated by the YSP, which contributes 63 -- 79 \% of the total flux in the normalizing bin.

With the inclusion of a power-law component (Figure \ref{q8ysppl} \& \ref{q2ysppl}), acceptable fits can be achieved for both an 8 Gyr OSP and a 2 Gyr ISP combined with a YSP and again, there are two regions for each combination in which $\chi^2_{red} < 1$. The younger region has ages in the range 5 -- 7 Myr but closer inspection of the detailed fits to the Ca II K line allows us to disregard these younger ages because the model substantially over-predicts the strength of this absorption feature.

The inclusion of the power-law component, which in this case could be representative of either a scattered AGN component or the continuum associated with very young stellar population (VYSP), results in a wider range of ages that provide acceptable fits (Table \ref{fit_info}). The YSP/ ISP ages range between 0.02 and 0.4 Gyr with reddening $0 \le E(B-V) \le 0.5 $. In these cases, the YSP contributes up to 94 \% of the total flux, while the power-law component contributes a maximum of 42\%. Unsurprisingly, as the contribution to the flux from the power law increases, the age and reddening of the YSP/ISP components that provide acceptable fits also increases. At the oldest ages and highest reddening, the power-law would contribute $\sim 40\%$ of the total flux in the normalising bin, which if due to scattered light from the hidden quasar would be detectable as broad wings to H$\beta$.

 In order to determine a reasonable upper limit for a contribution from scattered quasar light, we have added Gaussian components to the model-subtracted data with values of broad H$\beta$ equivalent width and FWHM that are typical of  quasars (78\AA\ and 5000 km s$^{-1}$ respectively)\footnote{These values were derived from 'A Catalog of Quasar Properties from SDSS DR7'\citep{schneider10,shen11}, by selecting all objects that have $L_{OIII} > 10^{42}$ erg s$^{-1}$ and taking the median values of broad H$\beta$ equivalent width and FWHM.}. We find that, if a scattered quasar component were present, this would be clearly detectable if the quasar contributed 20\% or more of the total flux in the normalising bin. Taking this into account rules out YSPs older than 100 Myr when considering both a 2 and 8 Gyr underlying OSP/ISP. Although these findings suggest that there is unlikely to be a major contribution from a scattered quasar component (associated with older age YSP), our observations do not allow us to rule out a lower level scattered quasar contribution. Spectropolarimetry observations are required to accurately quantify the true level of any scattered light component.

\subsection{The companion galaxy nucleus}

Figure \ref{g1ysp} \& \ref{g0-5ysp} show the contours of $\chi^2_{red}$ for the combinations of components that produce acceptable fits for the nucleus of the companion galaxy. These clearly demonstrate that a much narrower range of YSP ages produces acceptable fits. In fact, it is apparent from Table \ref{fit_info} that it is only fits including ISPs of both 0.5 or 2 Gyr and a YSP component of 9 -- 10 Myr with $0 \le E(B-V) \le 0.1 $ that produce acceptable results. In both cases, the flux is dominated by the YSP component, which contributes between 65\% and 97 \% of the total flux in the normalizing bin. Note that the presence of a YSP of relatively young age is consistent with the detection of strong emission lines in this region with H {\sevensize II} region like line ratios \citep{villar11}. Indeed comparison of the equivalent width of the H$\beta$ line ($log$(W(H$\beta$)\AA\ )=1.27) with the instantaneous burst models of \citet{leitherer99} implies $t_{ysp} \sim 6$~Myr, although older ages are possible if the star formation history is more extended. However, this must be considered a lower limit because, if the population were younger than this, we would expect to detect the broad He {\sevensize II}$\lambda \lambda 4686, 5411$ features  associated with Wolf-Rayet stars, which is not the case \citep{leitherer99}.

\subsection{The tidal tail} 

The northern tidal tail can be fit with an underlying OSP of 8 Gyr or ISPs of 0.5, 1 and 2 Gyr, all combined with a YSP. Table \ref{fit_info} shows that for all except the fits with a 0.5 Gyr underlying population, we require a YSP of 20--40 Myr with reddening $0 \le E(B-V) \le 0.1 $, comparable with the YSP population required for the quasar host galaxy component. For an ISP of 0.5 Gyr, there a two possible regions in $\chi^2_{red}$ space that produce acceptable fits. These have YSP ages 1--6 Myr ($0 \le E(B-V) \le 0.3 $) and 9--30 Myr ($0 \le E(B-V) \le 0.1 $). Table \ref{fit_info} shows that, when considering fits with an 8 or 2 Gyr component, it is the YSP that dominates the flux in the aperture. For ISPs of 0.5 and 1 Gyr, the ISP flux makes a more significant contribution to the total flux. Although visual inspection of the fits does not allow for discrimination between the older and younger combinations, the equivalent width of the detected H$\beta$ line suggests a starburst age of $\sim 6$ Myr \citep{leitherer99}, favouring the 0.5 Gyr + 1-6 Myr population fits. However, all OSP/ISP plus YSP combinations that provide an acceptable fit to the data require a contribution from a YSP with low reddening.

\subsection{Stellar Masses}

The relationship between quasar host galaxies and quiescent galaxies remains unclear: are quasars hosted predominantly by massive ellipticals, or can they be hosted by galaxies across the stellar mass range? One approach to answering this question involves determining the total stellar masses of the host galaxies using broad-band photometry at a single wavelength. However, when doing so it is important to account for the mix of stellar populations present, because the assumptions made can have a significant impact on the result. 

The accuracy of our modelling allows us to calculate the masses of the stellar populations within each aperture, accounting for the OSP/ISP and YSP separately. In addition, by using the photometry from our Gemini GMOS-S imaging data we can account for the extended flux that is not contained within the slit. 

We have calculated the possible ranges of masses in each of the apertures separately by first assuming that the OSP makes the maximum acceptable contribution to the total flux in that aperture, and then repeating the process for the maximum acceptable YSP flux. We have done this because, although the YSP dominates the flux in all three apertures, it is the OSP, particularly when an 8 Gyr OSP is assumed, that will most likely dominate the mass of the system. Table \ref{stellar_masses} shows the stellar masses derived for the OSP and YSP components separately and also the total stellar mass in each spatial region. Clearly the total stellar mass of the system is dominated by the 8 Gyr OSP that is detected in the quasar component (Q). In the companion galaxy region (G), the maximum mass is contributed by a 0.5 Gyr underlying ISP component, and therefore, this part of the system only contributes $\sim 3-8\%$ of the total stellar mass.

\begin{table}
\caption{The stellar masses derived from each of the 5 kpc apertures. Column 1 gives the aperture, while columns 2 to 4 show the derived mass of the OSP, YSP and total mass respectively. All the values are given in units of $10^9~M_{\sun}$. The results shown in this table correspond to the minimum and maximum contribution to the flux from an OSP/ISP.  The mass estimates assume a Kroupa IMF}
  \label{stellar_masses}
  \begin{tabular}{l c c c}
  \hline
  Aperture	&	OSP/ISP  		& YSP  			&	Total  \\
  \hline		
   Q		&	31--54		& 0.84--1.9 	& 32--54\\	
   G		&	0.32--5.7	& 0.39--0.86 	& 1.1--6.0\\
   T		&   0.24--6.8 	& 0.20 --0.37 	& 0.61--7.1\\
   \hline
  Total Mass&												&													& 33 -- 67\\
	\hline
  \end{tabular}
\end{table}

The results in Table \ref{stellar_masses} include only the flux contained within the 1.5 arcsec slit, which corresponds to an area of $5 \times 6.7 $ kpc$^2$ per aperture in our assumed cosmology. In order to correct for this, we make use of our deep GMOS-S $r^\prime$ imaging \citep{bessiere12}, from which we have determined a magnitude of $r_{AB} = 18.40$, (corrected for Galactic extinction), within a 30 kpc aperture. From this, we find that 59\% of the flux at the central wavelength of the $r^{\prime}$ filter is accounted for by the three apertures in the slit. 

To calculate the total mass of the system we assume that starlight not accounted for by the apertures comprises the same mix of populations as apertures Q, G and T, in turn, both for a minimum and maximum contribution from an OSP. Therefore, the minimum mass is derived by assuming the minimum OSP contribution to the models for each aperture and assuming the same mix of stellar populations as in aperture G for the flux excluded from the slit. The maximum mass is determined by summing the masses found from the models assuming the largest possible contribution from an OSP/ISP, and that the flux excluded from the aperture can be modelled with the same mix of populations as aperture Q. This leads to a total stellar mass for the system in the range $3.7 \times 10^{10} M_{\sun} < M < 1.7\times 10^{11} M_{\sun}$ which, if we adopt the mass function of \citet{cole01} ($m_*= 1.4 \times 10^{11}$M$_{\sun}$), is around the characteristic mass ($0.3m_* - 1.2m_*$). 

A number of previous studies of the host galaxies of both radio-loud and radio-quiet quasars have concluded that they tend to be hosted by massive elliptical galaxies. Scaling the derived values to a Kroupa IMF for ease of comparison, \citet{dunlop03} and \citet{tadhunter11} found median stellar masses of $4.1m_*$ and $2.5m_*$ respectively for radio-loud quasars. \citet{dunlop03} also derived masses for a sample of radio-quiet quasars finding a slightly lower median stellar mass of $1.8m_*$. Our stellar mass calculations, which take into account the dominant contribution to the flux from the YSP, also show that J0025-10 will not evolve into a massive elliptical through \emph{this} merger\footnote{We note that although there is some cold molecular gas remaining in this system -- as evidenced by the CO observations -- its mass is not sufficient to significantly boost the total stellar mass, even if completely converted in to stars.}. In fact, the stellar mass derived for this system is consistent with the stellar masses derived for a sample of ULIRGs in the local Universe by \citet{rodriguez10} (median mass 0.5$m_*$).

\section{Discussion}

The results presented above make it clear that in all three regions of J0025-10 a YSP of age $t_{ysp}< 40$ Myr with low reddening ($0 \le E(B-V) \le 0.3 $) contributes a large proportion of the total flux. It is striking that we find similar ages and reddening for these YSPs across the entire system, suggesting that J0025-10 is currently undergoing a galaxy-wide burst of star formation. This finding is interesting in the context of evolutionary scenarios, such as that of \citet{sanders88}, in which galaxy mergers lead to ULIRGs which then eventually become observable quasars. Such models predict that, at the peak of star formation activity, associated with the time of the coalescence of the two BHs (e.g. \citealt{dimatteo05,springel05,hopkins06}), we should observe a dust enshrouded system containing a highly reddened YSP (i.e. a ULIRG), in which any quasar activity that may have been triggered is too deeply embedded for us to detect directly. It is only with the onset of AGN-related feedback processes, as the AGN begins to dominate, that the gas and dust are removed, resulting in the quenching of star formation activity. Thus, in the \cite{sanders88} scenario, when the quasar becomes directly detectable, it should be hosted by a system in which the two nuclei have coalesced and also have an ageing starburst population.

 This predicted delay between the merger-induced starburst and visible AGN activity has indeed been detected by some previous stellar population studies \citep{fernandes04,tadhunter05,davies07,holt07,wills08,wild10,canalizo13,ramos13} where delays of a few 100 Myr to a few Gyr have been found. However, this is clearly not the whole story as others \citep{heckman97,canalizo00,canalizo01,brotherton02,holt07,wills08,tadhunter11} find evidence that the starburst and AGN activity have been triggered quasi-simultaneously.

One of the primary reasons for these apparently discrepant results may be that the triggering mechanisms depends on AGN luminosity. Indeed, \citet{tadhunter11} show that the vast majority of the objects in their sample of powerful radio galaxies that have quasar-like luminosities ($L_{[OIII]} > 10^{35}~W$), display clear evidence of a significant contribution to their fluxes from YSPs with ages $t_{YSP} < 0.1$ Gyr. On the other hand, less luminous AGN tend to show evidence for older YSP $0.2 < t_ysp < 2$ Gyr \citep{tadhunter05,emonts05,wild10}.

Taking this luminosity dependence into consideration, if we concentrate on the quasar-like objects with the best spectroscopic data on the stellar populations in the host galaxies \citep{canalizo00,holt07,wills08,tadhunter11}, in general, ages $t_{ysp} < 0.1$ Gyr are typical in the nuclear regions of the host galaxies. One notable exception to this trend is the study of \citet{canalizo13}, where they find that, although 14/15 quasar host galaxies do have evidence for a significant YSP, they are older ($t_{ISP} \sim 1.2$ Gyr) than those found in the majority of other quasar-like systems.

When comparing the results of such modelling, it is important to note that the dating of stellar populations is, in most cases, complex and has many inherent degeneracies, making it difficult to find  unique solutions. In fact, there are very few cases of quasar-like systems in which the stellar populations have been accurately dated. Some of the few clear cut cases include the powerful radio galaxy 3C459 \citep{wills08} and the nearby quasar Mrk 231 \citep{rodriguez09,canalizo01} in which the populations are found to have ages $t_{ysp} < 0.1$ Gyr. In the latter case, the findings are supported by the detection of He {\sevensize I} absorption features which are associated with B-type stars  \citep{gonzalez99} which have a lifetime of 20 -- 80 Myr. The nearby type II quasar Mrk 447 \citep{heckman97} is clearly identified as having a YSP with an age of $t_{YSP} \sim 6$ Myr, supported by the presence of UV absorption features such as Si {\sevensize III}$\lambda1417$, which are attributed to the presence of late O and early B supergiants. Thus, thanks to the quality and spectral range of the data, J0025-10 is included in a small group of objects in which the starburst population can be reliably dated, with the clear detection of the strong Balmer absorption features, Balmer break (Figure \ref{spectra}), and the dominant contribution to the total flux by the YSP. We strongly emphasise that in the cases in which the stellar populations can be unambiguously fit, it is found that a significant proportion of the flux is attributable to a YSP $t_{ysp} < 0.1$ Gyr.

In light of this, it important to understand why studies such as \citet{canalizo13} find such markedly different results. One possible reason for this discrepancy is that the latter study does not take into account reddening of the YSP. Although J0025-10 appears to be an exceptional case in which the YSP are unreddened, spectroscopic studies of the YSP in local radio galaxies and ULIRGs emphasise the importance of taking into account reddening, especially for the youngest YSP components \citep{tadhunter05,rodriguez09,rodriguez10}. Failure to take into account the reddening may lead to the ages of the YSP being over estimated. Alternatively, the discrepancy may be related to the fact that some of the studies of quasar-like systems concentrate on the near-nuclear bulge regions, whereas the study of Canalizo \& Stockton considered apertures that are, by necessity, offset by several kpc from the bright quasar nuclei. However, in this context we note that some of the apertures that give young YSP ages in Mrk231 and J0025-10 are also significantly offset from the AGN nuclei.

Although J0025-10 is classified as a ULIRG \citep{villar13}, this particular object displays a number of traits that are contradictory to the evolutionary view of \citet{sanders88}. To begin with its two nuclei have not yet coalesced and are still separated by $\sim 5$ kpc. Despite this, J0025-10 already hosts a luminous quasar, clearly detectable in the optical narrow lines, whilst at the same time hosting a massive YSP of $t_{ysp}< 40$ Myr. This strongly implies that the AGN and star formation activity were triggered quasi-simultaneously. However, we cannot rule out that at some future point, an even more intense period of star formation and BH accretion will occur.

We further note that this system appears to have very little (if any) reddening affecting the YSP or narrow emission lines, a result supported by both continuum and emission line modelling. In other words, there is a lack of dust in the regions in which a significant proportion of the YSP currently reside. The obvious implication of this is that the dust in these regions has already been efficiently removed by some process, most likely outflows powered by the AGN or starburst. If this is the case, then this process has been triggered earlier in the merger process than simulations have thus far suggested. However, assuming that the cool dust emitting the far-IR luminosity of the system (which suggests a star formation rate of $190 \pm 52$ M$_{\sun}$ yr$^{-1}$; \citealt{villar13}) is indeed heated by young stars, a substantial fraction of the star formation activity must remain enshrouded in dust, although we do not detect this population in our optical spectra.

\section{Conclusions}

We have used Gemini GMOS-S long-slit data to determine the ages and masses of the stellar populations in apertures extracted from three distinct regions of this merging type II quasar system. We have found the following.

\begin{enumerate}
\item The quasar host galaxy component of the system is well fit with a combination of an unreddened 8 Gyr population combined with a YSP of between 20 and 40 Myr with reddenings in the range $0 \le E(B-V) \le 0.2 $.
\item The companion galaxy is well fit with an ISP of 0.5-1 Gyr with a YSP of between 6 and 10 Myr (taking into account the H$\beta$ line equivalent width) and reddenings in the range $0 \le E(B-V) \le 0.1 $.
\item The north eastern tail component is well fit by an OSP/ISP and a YSP of 20--40 Myr with reddenings in the range $0 \le E(B-V) \le 0.1 $. It is also well fit by an ISP of 0.5 Gyr with a YSP of 1 - 6 Myr. Though we cannot discriminate by the quality of the fit alone, the measured equivalent width of the H$\beta$ line favours this latter combination.

\item The total stellar mass of the system is $3.7 \times 10^{10} M_{\sun} <$ M$ < 1.7 \times 10^{11} M_{\sun}$ ($0.3m_* - 1.2m_*$), which is lower than the stellar masses previously found for most quasar host galaxies.

\end{enumerate}
The young ages of the stellar populations in all three regions of J0025-10 probed by our high quality spectra, unambiguously demonstrate that the luminous quasar-like nucleus in J0025-10 has been triggered at around the same time as a major episode of star formation immediately prior to the coalescence of the nuclei in a gas-rich merger. The low reddening deduced for all three apertures using the YSP continuum and the narrow line emission measurements clearly indicate that, even at this early stage of the merger, outflows driven either by the AGN or starburst have been effective in driving much of the dust and gas from the visible star formation regions. These results are fully consistent with the trend found by \citet{tadhunter11}, that luminous quasars host starburst populations with $t_{ysp}< 0.1$ Gyr.

\section{Acknowledgements}
PB acknowledges support in the form of an STFC Ph.D. studentship. CRA acknowledges the Estallidos group through project PN AYA2010-21887-C04.04 This work is based on observations obtained at the Gemini Observatory, which is operated by the Association of Universities for Research in Astronomy, Inc., under a cooperative agreement with the NSF on behalf of the Gemini partnership: the National Science Foundation (United States), the Science and Technology Facilities Council (United Kingdom), the National Research Council (Canada),CONICYT (Chile), the Australian Research Council (Australia), Minist\'{e}rio da Ci\^{e}ncia, Tecnologia e Inova\c{c}\H{a}o (Brazil) and Ministerio de Ciencia, Tecnolog\'{i}a e Innovaci\'{o}n Productiva Argentina). The Gemini programs under which the data were obtained are GS-2009B-Q-87 and GS-2011B-Q-42.This research has made use of the NASA/ IPAC Infrared Science Archive, which is operated by the Jet Propulsion Laboratory, California Institute of Technology, under contract with the National Aeronautics and Space Administration.

\bsp
\label{lastpage}

\begin{thebibliography}{}

\bibitem[\protect\citeauthoryear{{Bennert}, {Canalizo}, {Jungwiert},
  {Stockton}, {Schweizer}, {Peng} \& {Lacy}}{{Bennert}
  et~al.}{2008}]{bennert08}
{Bennert} N.,  {Canalizo} G.,  {Jungwiert} B.,  {Stockton} A.,  {Schweizer} F.,
   {Peng} C.~Y.,    {Lacy} M.,  2008, ApJ, 677, 846

\bibitem[\protect\citeauthoryear{{Bessiere}, {Tadhunter}, {Ramos Almeida} \&
  {Villar Mart{\'{\i}}n}}{{Bessiere} et~al.}{2012}]{bessiere12}
{Bessiere} P.~S.,  {Tadhunter} C.~N.,  {Ramos Almeida} C.,    {Villar
  Mart{\'{\i}}n} M.,  2012, MNRAS, 426, 276

\bibitem[\protect\citeauthoryear{{Brotherton}, {Grabelsky}, {Canalizo}, {van
  Breugel}, {Filippenko}, {Croom}, {Boyle} \& {Shanks}}{{Brotherton}
  et~al.}{2002}]{brotherton02}
{Brotherton} M.~S.,  {Grabelsky} M.,  {Canalizo} G.,  {van Breugel} W.,
  {Filippenko} A.~V.,  {Croom} S.,  {Boyle} B.,    {Shanks} T.,  2002, PASP,
  114, 593

\bibitem[\protect\citeauthoryear{{Calzetti}, {Armus}, {Bohlin}, {Kinney},
  {Koornneef} \& {Storchi-Bergmann}}{{Calzetti} et~al.}{2000}]{calzetti00}
{Calzetti} D.,  {Armus} L.,  {Bohlin} R.~C.,  {Kinney} A.~L.,  {Koornneef} J.,
    {Storchi-Bergmann} T.,  2000, ApJ, 533, 682

\bibitem[\protect\citeauthoryear{{Canalizo} \& {Stockton}}{{Canalizo} \&
  {Stockton}}{2000}]{canalizo00}
{Canalizo} G.,  {Stockton} A.,  2000, ApJ, 528, 201

\bibitem[\protect\citeauthoryear{{Canalizo} \& {Stockton}}{{Canalizo} \&
  {Stockton}}{2001}]{canalizo01}
{Canalizo} G.,  {Stockton} A.,  2001, ApJ, 555, 719

\bibitem[\protect\citeauthoryear{{Canalizo} \& {Stockton}}{{Canalizo} \&
  {Stockton}}{2013}]{canalizo13}
{Canalizo} G.,  {Stockton} A.,  2013, ArXiv e-prints

\bibitem[\protect\citeauthoryear{{Cardelli}, {Clayton} \& {Mathis}}{{Cardelli}
  et~al.}{1989}]{cardelli89}
{Cardelli} J.~A.,  {Clayton} G.~C.,    {Mathis} J.~S.,  1989, ApJ, 345, 245

\bibitem[\protect\citeauthoryear{{Cid Fernandes}, {Gonz{\'a}lez Delgado},
  {Schmitt}, {Storchi-Bergmann}, {Martins}, {P{\'e}rez}, {Heckman}, {Leitherer}
  \& {Schaerer}}{{Cid Fernandes} et~al.}{2004}]{fernandes04}
{Cid Fernandes} R.,  {Gonz{\'a}lez Delgado} R.~M.,  {Schmitt} H.,
  {Storchi-Bergmann} T.,  {Martins} L.~P.,  {P{\'e}rez} E.,  {Heckman} T.,
  {Leitherer} C.,    {Schaerer} D.,  2004, ApJ, 605, 105

\bibitem[\protect\citeauthoryear{{Cole}, {Norberg}, {Baugh}, {Frenk},
  {Bland-Hawthorn}, {Bridges}, {Cannon}, {Colless}, {Collins}, {Couch},
  {Cross}, {Dalton}, {De Propris}, {Driver} \& {Efstathiou}}{{Cole}
  et~al.}{2001}]{cole01}
{Cole} S.,  {Norberg} P.,  {Baugh} C.~M.,  {Frenk} C.~S.,  {Bland-Hawthorn} J.,
   {Bridges} T.,  {Cannon} R.,  {Colless} M.,  {Collins} C.,  {Couch} W.,
  {Cross} N.,  {Dalton} G.,  {De Propris} R.,  {Driver} S.~P.,    {Efstathiou}
  G.,  2001, MNRAS, 326, 255

\bibitem[\protect\citeauthoryear{{Davies}, {M{\"u}ller S{\'a}nchez}, {Genzel},
  {Tacconi}, {Hicks}, {Friedrich} \& {Sternberg}}{{Davies}
  et~al.}{2007}]{davies07}
{Davies} R.~I.,  {M{\"u}ller S{\'a}nchez} F.,  {Genzel} R.,  {Tacconi} L.~J.,
  {Hicks} E.~K.~S.,  {Friedrich} S.,    {Sternberg} A.,  2007, ApJ, 671, 1388

\bibitem[\protect\citeauthoryear{{di Matteo}, {Springel} \& {Hernquist}}{{di
  Matteo} et~al.}{2005}]{dimatteo05}
{di Matteo} T.,  {Springel} V.,    {Hernquist} L.,  2005, Nat, 433, 604

\bibitem[\protect\citeauthoryear{{Dickson}, {Tadhunter}, {Shaw}, {Clark} \&
  {Morganti}}{{Dickson} et~al.}{1995}]{dickson95}
{Dickson} R.,  {Tadhunter} C.,  {Shaw} M.,  {Clark} N.,    {Morganti} R.,
  1995, MNRAS, 273, L29

\bibitem[\protect\citeauthoryear{{Dunlop}, {McLure}, {Kukula}, {Baum}, {O'Dea}
  \& {Hughes}}{{Dunlop} et~al.}{2003}]{dunlop03}
{Dunlop} J.~S.,  {McLure} R.~J.,  {Kukula} M.~J.,  {Baum} S.~A.,  {O'Dea}
  C.~P.,    {Hughes} D.~H.,  2003, MNRAS, 340, 1095

\bibitem[\protect\citeauthoryear{{Emonts}, {Morganti}, {Tadhunter},
  {Oosterloo}, {Holt} \& {van der Hulst}}{{Emonts} et~al.}{2005}]{emonts05}
{Emonts} B.~H.~C.,  {Morganti} R.,  {Tadhunter} C.~N.,  {Oosterloo} T.~A.,
  {Holt} J.,    {van der Hulst} J.~M.,  2005, MNRAS, 362, 931

\bibitem[\protect\citeauthoryear{{Ferrarese} \& {Merritt}}{{Ferrarese} \&
  {Merritt}}{2000}]{ferrarese00}
{Ferrarese} L.,  {Merritt} D.,  2000, ApJL, 539, L9

\bibitem[\protect\citeauthoryear{{Gebhardt}, {Bender}, {Bower}, {Dressler},
  {Faber}, {Filippenko}, {Green}, {Grillmair}, {Ho}, {Kormendy}, {Lauer},
  {Magorrian}, {Pinkney}, {Richstone} \& {Tremaine}}{{Gebhardt}
  et~al.}{2000}]{gebhardt00}
{Gebhardt} K.,  {Bender} R.,  {Bower} G.,  {Dressler} A.,  {Faber} S.~M.,
  {Filippenko} A.~V.,  {Green} R.,  {Grillmair} C.,  {Ho} L.~C.,  {Kormendy}
  J.,  {Lauer} T.~R.,  {Magorrian} J.,  {Pinkney} J.,  {Richstone} D.,
  {Tremaine} S.,  2000, ApJL, 539, L13

\bibitem[\protect\citeauthoryear{{Gonz{\'a}lez Delgado}, {Leitherer} \&
  {Heckman}}{{Gonz{\'a}lez Delgado} et~al.}{1999}]{gonzalez99}
{Gonz{\'a}lez Delgado} R.~M.,  {Leitherer} C.,    {Heckman} T.~M.,  1999, ApJs,
  125, 489

\bibitem[\protect\citeauthoryear{{Heckman}, {Gonzalez-Delgado}, {Leitherer},
  {Meurer}, {Krolik}, {Wilson}, {Koratkar} \& {Kinney}}{{Heckman}
  et~al.}{1997}]{heckman97}
{Heckman} T.~M.,  {Gonzalez-Delgado} R.,  {Leitherer} C.,  {Meurer} G.~R.,
  {Krolik} J.,  {Wilson} A.~S.,  {Koratkar} A.,    {Kinney} A.,  1997, ApJ,
  482, 114

\bibitem[\protect\citeauthoryear{{Heckman}, {Smith}, {Baum}, {van Breugel},
  {Miley}, {Illingworth}, {Bothun} \& {Balick}}{{Heckman}
  et~al.}{1986}]{heckman86}
{Heckman} T.~M.,  {Smith} E.~P.,  {Baum} S.~A.,  {van Breugel} W.~J.~M.,
  {Miley} G.~K.,  {Illingworth} G.~D.,  {Bothun} G.~D.,    {Balick} B.,  1986,
  ApJ, 311, 526

\bibitem[\protect\citeauthoryear{{Holt}, {Tadhunter}, {Gonz{\'a}lez Delgado},
  {Inskip}, {Rodriguez Zaurin}, {Emonts}, {Morganti} \& {Wills}}{{Holt}
  et~al.}{2007}]{holt07}
{Holt} J.,  {Tadhunter} C.~N.,  {Gonz{\'a}lez Delgado} R.~M.,  {Inskip} K.~J.,
  {Rodriguez Zaurin} J.,  {Emonts} B.~H.~C.,  {Morganti} R.,    {Wills} K.~A.,
  2007, MNRAS, 381, 611

\bibitem[\protect\citeauthoryear{{Holt}, {Tadhunter} \& {Morganti}}{{Holt}
  et~al.}{2003}]{holt03}
{Holt} J.,  {Tadhunter} C.~N.,    {Morganti} R.,  2003, MNRAS, 342, 227

\bibitem[\protect\citeauthoryear{{Hopkins}, {Hernquist}, {Cox}, {Di Matteo},
  {Robertson} \& {Springel}}{{Hopkins} et~al.}{2006}]{hopkins06}
{Hopkins} P.~F.,  {Hernquist} L.,  {Cox} T.~J.,  {Di Matteo} T.,  {Robertson}
  B.,    {Springel} V.,  2006, ApJS, 163, 1

\bibitem[\protect\citeauthoryear{{Kroupa}}{{Kroupa}}{2001}]{kroupa01}
{Kroupa} P.,  2001, MNRAS, 322, 231

\bibitem[\protect\citeauthoryear{{Leitherer}, {Ortiz Ot{\'a}lvaro}, {Bresolin},
  {Kudritzki}, {Lo Faro}, {Pauldrach}, {Pettini} \& {Rix}}{{Leitherer}
  et~al.}{2010}]{leitherer10}
{Leitherer} C.,  {Ortiz Ot{\'a}lvaro} P.~A.,  {Bresolin} F.,  {Kudritzki}
  R.-P.,  {Lo Faro} B.,  {Pauldrach} A.~W.~A.,  {Pettini} M.,    {Rix} S.~A.,
  2010, ApJS, 189, 309

\bibitem[\protect\citeauthoryear{{Leitherer}, {Schaerer}, {Goldader},
  {Gonz{\'a}lez Delgado}, {Robert}, {Kune}, {de Mello}, {Devost} \&
  {Heckman}}{{Leitherer} et~al.}{1999}]{leitherer99}
{Leitherer} C.,  {Schaerer} D.,  {Goldader} J.~D.,  {Gonz{\'a}lez Delgado}
  R.~M.,  {Robert} C.,  {Kune} D.~F.,  {de Mello} D.~F.,  {Devost} D.,
  {Heckman} T.~M.,  1999, ApJS, 123, 3

\bibitem[\protect\citeauthoryear{{Magorrian}, {Tremaine}, {Richstone},
  {Bender}, {Bower}, {Dressler}, {Faber}, {Gebhardt}, {Green}, {Grillmair},
  {Kormendy} \& {Lauer}}{{Magorrian} et~al.}{1998}]{magorrian98}
{Magorrian} J.,  {Tremaine} S.,  {Richstone} D.,  {Bender} R.,  {Bower} G.,
  {Dressler} A.,  {Faber} S.~M.,  {Gebhardt} K.,  {Green} R.,  {Grillmair} C.,
  {Kormendy} J.,    {Lauer} T.,  1998, AJ, 115, 2285

\bibitem[\protect\citeauthoryear{{Martini} \& {Weinberg}}{{Martini} \&
  {Weinberg}}{2001}]{martini01}
{Martini} P.,  {Weinberg} D.~H.,  2001, ApJ, 547, 12

\bibitem[\protect\citeauthoryear{{Osterbrock} \& {Ferland}}{{Osterbrock} \&
  {Ferland}}{2006}]{osterbrock06}
{Osterbrock} D.~E.,  {Ferland} G.~J.,  2006, {Astrophysics of gaseous nebulae
  and active galactic nuclei}

\bibitem[\protect\citeauthoryear{{Ramos Almeida}, {Bessiere}, {Tadhunter},
  {P{\'e}rez-Gonz{\'a}lez}, {Barro}, {Inskip}, {Morganti}, {Holt} \&
  {Dicken}}{{Ramos Almeida} et~al.}{2012}]{ramos12}
{Ramos Almeida} C.,  {Bessiere} P.~S.,  {Tadhunter} C.~N.,
  {P{\'e}rez-Gonz{\'a}lez} P.~G.,  {Barro} G.,  {Inskip} K.~J.,  {Morganti} R.,
   {Holt} J.,    {Dicken} D.,  2012, MNRAS, 419, 687

\bibitem[\protect\citeauthoryear{{Ramos Almeida}, {Rodr{\'{\i}}guez Espinosa},
  {Acosta-Pulido}, {Alonso-Herrero}, {P{\'e}rez Garc{\'{\i}}a} \&
  {Rodr{\'{\i}}guez-Eugenio}}{{Ramos Almeida} et~al.}{2013}]{ramos13}
{Ramos Almeida} C.,  {Rodr{\'{\i}}guez Espinosa} J.~M.,  {Acosta-Pulido} J.~A.,
   {Alonso-Herrero} A.,  {P{\'e}rez Garc{\'{\i}}a} A.~M.,
  {Rodr{\'{\i}}guez-Eugenio} N.,  2013, MNRAS, 429, 3449

\bibitem[\protect\citeauthoryear{{Ramos Almeida}, {Tadhunter}, {Inskip},
  {Morganti}, {Holt} \& {Dicken}}{{Ramos Almeida} et~al.}{2011}]{ramos11}
{Ramos Almeida} C.,  {Tadhunter} C.~N.,  {Inskip} K.~J.,  {Morganti} R.,
  {Holt} J.,    {Dicken} D.,  2011, MNRAS, 410, 1550

\bibitem[\protect\citeauthoryear{{Robinson}}{{Robinson}}{2001}]{robinson01}
{Robinson} T.,  2001, PhD thesis, Univ. Sheffield

\bibitem[\protect\citeauthoryear{{Robinson}, {Tadhunter}, {Axon} \&
  {Robinson}}{{Robinson} et~al.}{2000}]{robinson00}
{Robinson} T.~G.,  {Tadhunter} C.~N.,  {Axon} D.~J.,    {Robinson} A.,  2000,
  MNRAS, 317, 922

\bibitem[\protect\citeauthoryear{{Rodr{\'{\i}}guez Zaur{\'{\i}}n}, {Tadhunter}
  \& {Gonz{\'a}lez Delgado}}{{Rodr{\'{\i}}guez Zaur{\'{\i}}n}
  et~al.}{2009}]{rodriguez09}
{Rodr{\'{\i}}guez Zaur{\'{\i}}n} J.,  {Tadhunter} C.~N.,    {Gonz{\'a}lez
  Delgado} R.~M.,  2009, MNRAS, 400, 1139

\bibitem[\protect\citeauthoryear{{Rodr{\'{\i}}guez Zaur{\'{\i}}n}, {Tadhunter}
  \& {Gonz{\'a}lez Delgado}}{{Rodr{\'{\i}}guez Zaur{\'{\i}}n}
  et~al.}{2010}]{rodriguez10}
{Rodr{\'{\i}}guez Zaur{\'{\i}}n} J.,  {Tadhunter} C.~N.,    {Gonz{\'a}lez
  Delgado} R.~M.,  2010, MNRAS, 403, 1317

\bibitem[\protect\citeauthoryear{{Sanders}, {Soifer}, {Elias}, {Madore},
  {Matthews}, {Neugebauer} \& {Scoville}}{{Sanders} et~al.}{1988}]{sanders88}
{Sanders} D.~B.,  {Soifer} B.~T.,  {Elias} J.~H.,  {Madore} B.~F.,  {Matthews}
  K.,  {Neugebauer} G.,    {Scoville} N.~Z.,  1988, ApJ, 325, 74

\bibitem[\protect\citeauthoryear{{Schlegel}, {Finkbeiner} \&
  {Davis}}{{Schlegel} et~al.}{1998}]{schlegel98}
{Schlegel} D.~J.,  {Finkbeiner} D.~P.,    {Davis} M.,  1998, ApJ, 500, 525

\bibitem[\protect\citeauthoryear{{Schneider}, {Richards}, {Hall}, {Strauss},
  {Anderson}, {Boroson}, {Ross}, {Shen}, {Brandt}, {Fan}, {Inada}, {Jester},
  {Knapp}, {Krawczyk}, {Thakar}, {Vanden Berk}, {Voges}, {Yanny} \&
  {York}}{{Schneider} et~al.}{2010}]{schneider10}
{Schneider} D.~P.,  et al. ,  2010, AJ, 139, 2360

\bibitem[\protect\citeauthoryear{{Shen}, {Richards}, {Strauss}, {Hall},
  {Schneider}, {Snedden}, {Bizyaev}, {Brewington}, {Malanushenko},
  {Malanushenko}, {Oravetz}, {Pan} \& {Simmons}}{{Shen} et~al.}{2011}]{shen11}
{Shen} Y.,  {Richards} G.~T.,  {Strauss} M.~A.,  {Hall} P.~B.,  {Schneider}
  D.~P.,  {Snedden} S.,  {Bizyaev} D.,  {Brewington} H.,  {Malanushenko} V.,
  {Malanushenko} E.,  {Oravetz} D.,  {Pan} K.,    {Simmons} A.,  2011, ApJS,
  194, 45

\bibitem[\protect\citeauthoryear{{Springel}, {Di Matteo} \&
  {Hernquist}}{{Springel} et~al.}{2005}]{springel05}
{Springel} V.,  {Di Matteo} T.,    {Hernquist} L.,  2005, MNRAS, 361, 776

\bibitem[\protect\citeauthoryear{{Tadhunter}, {Dickson}, {Morganti},
  {Robinson}, {Wills}, {Villar-Martin} \& {Hughes}}{{Tadhunter}
  et~al.}{2002}]{tadhunter02}
{Tadhunter} C.,  {Dickson} R.,  {Morganti} R.,  {Robinson} T.~G.,  {Wills} K.,
  {Villar-Martin} M.,    {Hughes} M.,  2002, MNRAS, 330, 977

\bibitem[\protect\citeauthoryear{{Tadhunter}, {Holt}, {Gonz{\'a}lez Delgado},
  {Rodr{\'{\i}}guez Zaur{\'{\i}}n}, {Villar-Mart{\'{\i}}n}, {Morganti},
  {Emonts}, {Ramos Almeida} \& {Inskip}}{{Tadhunter}
  et~al.}{2011}]{tadhunter11}
{Tadhunter} C.,  {Holt} J.,  {Gonz{\'a}lez Delgado} R.,  {Rodr{\'{\i}}guez
  Zaur{\'{\i}}n} J.,  {Villar-Mart{\'{\i}}n} M.,  {Morganti} R.,  {Emonts} B.,
  {Ramos Almeida} C.,    {Inskip} K.,  2011, MNRAS, 412, 960

\bibitem[\protect\citeauthoryear{{Tadhunter}, {Robinson}, {Gonz{\'a}lez
  Delgado}, {Wills} \& {Morganti}}{{Tadhunter} et~al.}{2005}]{tadhunter05}
{Tadhunter} C.,  {Robinson} T.~G.,  {Gonz{\'a}lez Delgado} R.~M.,  {Wills} K.,
    {Morganti} R.,  2005, MNRAS, 356, 480

\bibitem[\protect\citeauthoryear{{Toomre} \& {Toomre}}{{Toomre} \&
  {Toomre}}{1972}]{toomre72}
{Toomre} A.,  {Toomre} J.,  1972, ApJ, 178, 623

\bibitem[\protect\citeauthoryear{{V{\'a}zquez} \& {Leitherer}}{{V{\'a}zquez} \&
  {Leitherer}}{2005}]{vazquez05}
{V{\'a}zquez} G.~A.,  {Leitherer} C.,  2005, ApJ, 621, 695

\bibitem[\protect\citeauthoryear{{Villar-Mart{\'{\i}}n}, {Cabrera Lavers},
  {Bessiere}, {Tadhunter}, {Rose} \& {de Breuck}}{{Villar-Mart{\'{\i}}n}
  et~al.}{2012}]{villar12}
{Villar-Mart{\'{\i}}n} M.,  {Cabrera Lavers} A.,  {Bessiere} P.,  {Tadhunter}
  C.,  {Rose} M.,    {de Breuck} C.,  2012, MNRAS, 423, 80

\bibitem[\protect\citeauthoryear{{Villar-Martin}, {Emonts}, {Rodriguez}, {Perez
  Torres} \& {Drouart}}{{Villar-Martin} et~al.}{2013}]{villar13}
{Villar-Martin} M.,  {Emonts} B.,  {Rodriguez} M.,  {Perez Torres} M.,
  {Drouart} G.,  2013, ArXiv e-prints

\bibitem[\protect\citeauthoryear{{Villar-Mart{\'{\i}}n}, {Tadhunter},
  {Humphrey}, {Encina}, {Delgado}, {Torres} \&
  {Mart{\'{\i}}nez-Sansigre}}{{Villar-Mart{\'{\i}}n} et~al.}{2011}]{villar11}
{Villar-Mart{\'{\i}}n} M.,  {Tadhunter} C.,  {Humphrey} A.,  {Encina} R.~F.,
  {Delgado} R.~G.,  {Torres} M.~P.,    {Mart{\'{\i}}nez-Sansigre} A.,  2011,
  MNRAS, 416, 262

\bibitem[\protect\citeauthoryear{{Wild}, {Heckman} \& {Charlot}}{{Wild}
  et~al.}{2010}]{wild10}
{Wild} V.,  {Heckman} T.,    {Charlot} S.,  2010, MNRAS, 405, 933

\bibitem[\protect\citeauthoryear{{Wills}, {Tadhunter}, {Holt}, {Gonz{\'a}lez
  Delgado}, {Inskip}, {Rodr{\'{\i}}guez Zaur{\'{\i}}n} \& {Morganti}}{{Wills}
  et~al.}{2008}]{wills08}
{Wills} K.~A.,  {Tadhunter} C.,  {Holt} J.,  {Gonz{\'a}lez Delgado} R.,
  {Inskip} K.~J.,  {Rodr{\'{\i}}guez Zaur{\'{\i}}n} J.,    {Morganti} R.,
  2008, MNRAS, 385, 136

\bibitem[\protect\citeauthoryear{{Zakamska}, {Strauss}, {Krolik}, {Collinge},
  {Hall}, {Hao}, {Heckman}, {Ivezi{\'c}}, {Richards}, {Schlegel}, {Schneider},
  {Strateva}, {Vanden Berk}, {Anderson} \& {Brinkmann}}{{Zakamska}
  et~al.}{2003}]{zakamska03}
{Zakamska} N.~L.,  {Strauss} M.~A.,  {Krolik} J.~H.,  {Collinge} M.~J.,  {Hall}
  P.~B.,  {Hao} L.,  {Heckman} T.~M.,  {Ivezi{\'c}} {\v Z}.,  {Richards} G.~T.,
   {Schlegel} D.~J.,  {Schneider} D.~P.,  {Strateva} I.,  {Vanden Berk} D.~E.,
  {Anderson} S.~F.,    {Brinkmann} J.,  2003, AJ, 126, 2125

\end{thebibliography}
\end{document}